\documentclass[aip,amsmath,amssymb,reprint,onecolumn,nofootinbib]{revtex4-2}
\pdfoutput=1
\usepackage{physics2,derivative,cancel,color,graphicx,hyperref}
\usepackage{tikz}
\usetikzlibrary{decorations.pathmorphing}
\usetikzlibrary{decorations.markings}
\usephysicsmodule{ab,ab.braket}
\DeclareMathOperator{\Ai}{Ai}
\DeclareMathOperator{\Bi}{Bi}

\DeclareMathOperator{\arccosh}{arccosh}
\DeclareMathOperator{\order}{\mathcal O}

\begin{document}
\count\footins = 1000
\title{Topological Blocking of the Schwinger Effect in the Salpeter Equation: A Lefschetz Thimble Analysis}

\numberwithin{equation}{section}

\allowdisplaybreaks[1]

\author{Yutaro Shoji}
\affiliation{Jo\v{z}ef Stefan Institute, Jamova 39, 1000 Ljubljana, Slovenia}
\affiliation{Centre For Cosmology and Science Popularization (CCSP), SGT University, Gurugram, Delhi-NCR, Haryana 122505, India}

\begin{abstract}
We present a comprehensive Lefschetz thimble analysis of the one-dimensional Salpeter equation under a strong electric field.
By treating the non-local square-root operator within the framework of algebraic analysis, we construct the full solution space, which includes relativistic generalizations of the Airy $\Ai$ and $\Bi$ functions and their negative-energy counterparts. Through a direct comparison with the Dirac and Klein-Gordon equations, we provide a geometric explanation for the absence of Klein paradox and the Schwinger effect in the Salpeter equation.
Furthermore, our findings establish a unified geometric interpretation of the Schwinger effect across different relativistic wave equations.
\end{abstract}


\maketitle
\section{Introduction}
The fundamental equations of relativistic quantum mechanics are derived by elevating the relativistic energy-momentum relation $E^2 = p^2+m^2$ to an operator acting on wave functions. A direct generalization of the Schr\"odinger equation, achieved by taking the square root of this relation, yields the spinless Salpeter equation: $idf/dt  = \sqrt{-d^2/dx^2 + m^2} f$. This equation has been used as an approximation to the Bethe-Salpeter equation \cite{PhysRev.87.328}, and naturally arises in describing topological defects in Ising models \cite{PhysRevB.60.14525} and mesons in nuclear physics \cite{Allen:2003wz,Buisseret:2006sz}.
However, its Lorentz covariance is not manifest, and the non-local nature of the differential operator poses significant challenges for obtaining solutions.
Other relativistic equations, such as the Dirac equation and the Klein-Gordon equation, exhibit manifest Lorentz covariance and are more amenable to quantum field theory: The Dirac equation employs Dirac matrices to linearize the relativistic energy-momentum relation, while the Klein-Gordon equation applies the squared operator directly to the wave function.
Although their classical behaviors are similar \cite{10.1063/1.530015}, their quantum behaviors diverge significantly in the regime where $p^2$ is negative.

An example of this divergence is found in the Klein paradox (or Klein tunneling) \cite{1929ZPhy...53..157K,10.1119/1.1934851} and the Schwinger effect \cite{PhysRev.82.664,Sauter:1931zz,Heisenberg:1936nmg,Weisskopf:406571,Hansen_1981}, which involve tunneling between positive and negative energy branches.
With a linear scalar potential $V(x) = g x$, where $g > 0$, the classical momentum $p = \pm \sqrt{(E - V(x))^2 - m^2}$ becomes real in the regions $x < (E - m)/g$ and $x > (E + m)/g$, while it is purely imaginary between these regions. The Klein paradox refers to the counterintuitive behaviors of the transmission coefficient in this tunneling process: it increases as the potential becomes steeper for the Dirac equation and is even negative for the Klein-Gordon equation. These behaviors are interpreted as pair creation of particles and antiparticles around the potential barrier in the framework of second-quantized theory, which is referred to as the Schwinger effect.
The difference in transmission coefficients turns out to be essential for maintaining consistency with the commutation and anti-commutation relations of the creation and annihilation operators.

Conversely, it has been known that the Salpeter equation exhibits no Klein tunneling \cite{Daem:2024xrc,Zumer:2025jqh}, which is often attributed to its restriction to positive energies.
We contend, however, that within the framework of algebraic analysis, the square-root operator is more properly interpreted as a general symbol acting on a cycle and a kernel, thereby removing this restriction. Indeed, our results show that general solutions to the Salpeter equation encompass both positive and negative energy components.
It is therefore imperative to identify a more fundamental explanation for the absence of Klein tunneling in the Salpeter equation.

We present a comprehensive Lefschetz thimble analysis of the time-independent one-dimensional Salpeter equation with a linear potential, which is also known as the relativistic Airy equation. By comparing with the same analysis for the Dirac and Klein-Gordon equations, we find that the absence of Klein tunneling is due to the avoidance of Lefschetz thimbles from divergent branch points, which are not present in the Dirac and Klein-Gordon equations.
The validity of this explanation extends to complex parameters, a regime in which the functions may no longer be tempered distributions.
Therefore, this argument is topologically protected, providing a more fundamental explanation.
Furthermore, our approach provides a unified framework for understanding Klein tunneling in relativistic wave equations through Lefschetz thimbles.

From a mathematical perspective, our analysis carries implications for the Riemann-Hilbert correspondence \cite{Kashiwara1984,CM_1984__51_1_51_0}, which provides the formal foundation for Lefschetz thimble analysis \cite{Witten:2010cx}.
Recent advancements include formal proofs of the correspondence for irregular holonomic $\mathcal{D}$-modules \cite{2019arXiv190801276D} and regular holonomic $\mathcal{E}$-modules \cite{Waschkies2005MicrolocalRC}. While $\mathcal{D}$-modules involve polynomial differential operators, $\mathcal{E}$-modules accommodate pseudo-differential operators such as the square-root operator. The distinction between regular and irregular cases typically characterizes the asymptotic behavior of solutions as $|x|\to\infty$.
The Salpeter operator belongs to the class of irregular holonomic $\mathcal{E}$-modules, for which a rigorous proof of the correspondence remains an open question. Nonetheless, the correspondence is widely accepted, and it can be directly verified that the pairings of kernels and cycles yield valid solutions to the Salpeter equation. A remaining challenge is to establish that these pairings exhaust the full solution space.
We demonstrate that the correspondence successfully recovers both $\Ai$-type and $\Bi$-type solutions for the Salpeter equation, along with their negative-energy counterparts.
This recovery is consistent with the non-relativistic limit and suggests the applicability of the correspondence to this system, although the rigorous proof of the correspondence remains open.

The remainder of this paper is organized as follows. Section~\ref{sec:squareroot} details the treatment of the non-local square-root operator within the framework of algebraic analysis. Sections~\ref{sec:salpeter} and \ref{sec:dirac} present the Lefschetz thimble analysis for the Salpeter, Dirac, and Klein-Gordon equations. Utilizing these results, Section~\ref{sec:schwinger} discusses the Klein paradox and the Schwinger effect. Finally, we summarize in Section~\ref{sec:summary}.
\section{Non-local differential operator}\label{sec:squareroot}
The Salpeter equation involves a non-local differential operator that requires careful mathematical treatment.
In particular, conventional Fourier space techniques yield only a single solution corresponding to the Airy $\Ai$ function, which belongs to the tempered distribution space $\mathcal S'(\mathbb R)$.
However, a complete description of the solution space must accommodate the Airy $\Bi$ function, which lies outside $\mathcal S'(\mathbb R)$ and thus cannot be obtained through standard methods.
To systematically construct the full solution space, we employ the Riemann-Hilbert correspondence, which we detail below.

Let us start with a function $f$ belonging to the Schwartz space $\mathcal S(\mathbb R)$. In the conventional Fourier space techniques, the action of the operator is defined as
\begin{equation}
    \sqrt{-\partial^2+m^2}f(x)=\int_{-\infty}^\infty\frac{\odif{p}}{2\pi}\sqrt{p^2+m^2}\tilde f(p)e^{ipx},
\end{equation}
where $\partial$ denotes differentiation with respect to $x$ and $\tilde f$ is the Fourier transform of $f$.
However, this approach has inherent limitations, as the functions are restricted to the Schwartz space, meaning that their Fourier transforms must be well-defined. This often precludes the construction of general solutions to differential equations. The framework of microlocal analysis extends the application of a non-local operator to more general functions by reformulating the momentum integral as an integral over a cycle in the complex plane. The Riemann-Hilbert correspondence then conjecturally ensures that all solutions can be constructed by identifying all pairings of integral kernels $e^{\mathcal I_j(x,p)}$ and fast-decay homology cycles $\gamma$. Here, each pair yields a function,
\begin{equation}
    f_{j,\gamma}(x)=\int_{\gamma}\frac{\odif{p}}{2\pi}e^{\mathcal I_j(x,p)},\label{eq:ansatz}
\end{equation}
where $\mathcal{I}_j(x,p)$ is generally a multi-valued function of $p$, but its value must be smooth along the cycle $\gamma$.
When $\gamma$ lies on the real axis, this reproduces the standard Fourier transform, thereby encompassing functions in $\mathcal{S}(\mathbb{R})$.
This framework does not necessitate that solutions belong to $\mathcal{S}(\mathbb{R})$ or $\mathcal{S}'(\mathbb{R})$ since the cycle is always chosen such that the integral converges.

For the square-root operator, the preceding argument requires generalization, particularly concerning the space of $p$.
Consider a generalized action of the operator on the plane wave $f_p(x) = e^{ipx}$ for $p \in \mathbb{C}$, which may be defined such that $\sqrt{-\partial^2 + m^2} f_p(x) = \sqrt{p^2 + m^2} f_p(x)$.
However, this introduces a subtlety at the branch cut, where the sign of the operator becomes ambiguous, as $\lim_{\Re p \to 0^+} \sqrt{-\partial^2 + m^2} f_p(x) = -\lim_{\Re p \to 0^-} \sqrt{-\partial^2 + m^2} f_p(x)$ for $\Im(p) > m$, assuming the standard branch cut.
This indicates that the action of the operator is intrinsically defined on the Riemann surface,
\begin{equation}
    \Sigma=\{(p,T)\in \mathbb C^2\mid T^2=m^2+p^2\}.
\end{equation}
We refer to the branch with $T=+\sqrt{m^2+p^2}$ as the positive Riemann sheet, and the branch with $T=-\sqrt{m^2+p^2}$ as the negative Riemann sheet.
For a kernel $\mathcal{I}_j(x,p)$ characterized by a linear $x$-dependence of the form $ipx$, we define the action of the operator as
\begin{equation}
    \sqrt{-\partial^2+m^2}f_{j,\gamma}(x)=\int_{\gamma}\frac{\odif{p}}{2\pi}Te^{\mathcal I_j(x,p)},
\end{equation}
where $\gamma$ and $\odif{p}$ are a fast-decay cycle and a 1-form on $\Sigma$, respectively.
Notice that the operator acts on the whole domain $(x\in\mathbb C)$ of the function simultaneously due to the non-local nature of the operator.

The following example illustrates how this treatment addresses the two branches of the square-root operator. For all $q \in \mathbb{C}$, taking $\mathcal{I}_j(x,p) = ipx - \ln(p - q) - i\pi/2$ and $\gamma$ as a cycle enclosing $p = q$ on the positive Riemann sheet yields $f(x) = e^{iqx}$ and $\sqrt{-\partial^2 + m^2} f(x) = \sqrt{q^2 + m^2} f(x)$. Conversely, if $\gamma$ encloses $p = q$ on the negative Riemann sheet, we obtain $\sqrt{-\partial^2 + m^2} f(x) = -\sqrt{q^2 + m^2} f(x)$, despite $f(x)$ being identical.
This observation indicates that the operator does not act on the space of functions alone but on objects with richer structure. This aligns with the treatment of differential equations in algebraic analysis, as represented by the theory of $\mathcal{D}$-modules. The relevant structure comprises a pair of a kernel and a cycle, with analytic functions emerging as shadows generated by integrating the kernel over the cycle. Consequently, the resulting functions do not always retain complete information about the operator's action. This limitation motivates efforts to recover the information of pertinent cycles and kernels from analytic functions, a central theme in resurgence \cite{Ecalle1981LesFR} and exact WKB methods \cite{Voros1983TheRO,Kawai2005AlgebraicAO}.

\section{Lefschetz thimble analysis on Salpeter equation}\label{sec:salpeter}
In this section, we give a Lefschetz thimble analysis of the one-dimensional time-independent Salpeter equation with a linear potential, also known as the relativistic Airy equation:
\begin{equation}
    \ab[\sqrt{-\partial^2+m^2}+gx]f(x)=0,\label{eq:salpeter}
\end{equation}
where $m$ and $g$ are nonzero complex parameters. In physical applications, $m$ is the particle mass and $g$ is understood as the product of the particle charge $e$ and the electric field strength $\mathcal E$.
In general, one could consider a non-zero energy eigenvalue $E$; however, this merely corresponds to a shift in the spatial coordinate via $x\to x-E/g$.
\subsection{Kernel for Salpeter equation}
We employ the Ansatz given in Eq.~\eqref{eq:ansatz} with $\mathcal I_S(x,p)=ipx+iS_S(p)$. Substituting this into the Salpeter equation yields
\begin{align}
    0=\ab[\sqrt{-\partial^2+m^2}+gx]\int_{\gamma}\frac{\odif{p}}{2\pi}e^{\mathcal I_S(x,p)}&=\int_{\gamma}\frac{\odif{p}}{2\pi}e^{ipx+iS_S(p)}\ab(T-g\odv{S_S}{p}).
\end{align}
Here, integration by parts produces no boundary contributions because of the fast-decay nature of the cycle. The kernel is unique, and its exponent $S_S$ is obtained by integrating the 1-form $\eta=T\,\mathrm{d}p/g$ on $\Sigma$. Here, $\eta$ has two third-order poles at $(p,T)=(\infty,\pm\infty)$. Since $\Sigma$ is a genus-zero Riemann surface, it is topologically a sphere. Removing the two poles from $\Sigma$ yields a cylinder, which admits a single nontrivial homology cycle. Consequently, the integral of $\eta$ is not fixed solely by the endpoints of the cycle, but is indexed by an integer winding number $n$.

\begin{figure}
    \centering
    \begin{tikzpicture}
        \begin{scope}
        \clip (-3,-2) -- (3,-2) -- (3,2) -- (-3,2) -- cycle;
        \fill[orange!60] plot[domain=-3:-0.1] (\x, {2/\x}) -- (-3,-3) -- cycle;
        \fill[orange!60] plot[domain=0.1:3] (\x, {-2/\x}) -- (3,-3) -- cycle;
        \fill[blue!60] plot[domain=-3:-0.1] (\x, {-2/\x}) -- (-3,3) -- cycle;
        \fill[blue!60] plot[domain=0.1:3] (\x, {2/\x}) -- (3,3) -- cycle;
        \draw[thick, magenta, style={decoration={markings, mark=at position 0.6 with {\arrow{>}}},postaction={decorate}}] plot[domain=-3:3] (\x, {0.1*\x*\x+0.5});
        \draw[thick, green, style={decoration={markings, mark=at position 0.6 with {\arrow{>}}},postaction={decorate}}] plot[domain=-3:0] (\x, {-0.9*\x-1.5});
        \draw[thick, dashed, green] plot[domain=0:3] (\x, {-0.3*\x-1.5});
        \draw[thick, green, style={decoration={markings, mark=at position 0.6 with {\arrow{>}}},postaction={decorate}}] plot[domain=3:0] (\x, {0.9*\x-1.5});
        \draw[thick, dashed, green] plot[domain=0:-3] (\x, {0.3*\x-1.5});
        \end{scope}
        \draw[->, thick] (-3.1,0) -- (3.1,0) node[right] {$\Re(p)$};
        \draw[->, thick] (0,-2.1) -- (0,2.1) node[above] {$\Im(p)$};
        \draw[thick, decorate, decoration = {snake,amplitude=2pt,segment length=5pt}] (0,1) -- (0,2);
        \draw[thick, decorate, decoration = {snake,amplitude=2pt,segment length=5pt}] (0,-1) -- (0,-2);
        \filldraw[black] (0,1) circle (2pt) node[above right] {$im$};
        \filldraw[black] (0,-1) circle (2pt) node[below right] {$-im$};
        \node at (-2.5,1.7) {$-$};
        \node at (2.5,1.7) {$-$};
        \node at (2.5,-1.7) {$+$};
        \node at (-2.5,-1.7) {$+$};
        \node at (-0.6,0.7) {$\gamma_{\Ai}$};
        \node at (-1.6,-0.7) {$\gamma_{\Bi}^{(1)}$};
        \node at (1.6,-0.7) {$\gamma_{\Bi}^{(2)}$};
    \end{tikzpicture}
    \quad
    \begin{tikzpicture}
        \begin{scope}
        \clip (-3,-2) -- (3,-2) -- (3,2) -- (-3,2) -- cycle;
        \fill[orange!60] (3,1.35) circle (0.45);
        \fill[orange!60] (-3,1.35) circle (0.45);
        \fill[blue!60] (3,0.45) circle (0.45);
        \fill[blue!60] (-3,0.45) circle (0.45);
        \fill[blue!60] (3,-1.35) circle (0.45);
        \fill[blue!60] (-3,-1.35) circle (0.45);
        \fill[orange!60] (3,-0.45) circle (0.45);
        \fill[orange!60] (-3,-0.45) circle (0.45);
        \fill[blue!60] (3,2.25) circle (0.45);
        \fill[blue!60] (-3,2.25) circle (0.45);
        \fill[orange!60] (3,-2.25) circle (0.45);
        \fill[orange!60] (-3,-2.25) circle (0.45);
        \draw[thick, magenta, style={decoration={markings, mark=at position 0.65 with {\arrow{>}}},postaction={decorate}}] (-3,0.3) -- (3,0.3);
        \draw[thick, green, style={decoration={markings, mark=at position 0. with {\arrow{<}}},postaction={decorate}}] (-3,-0.45) circle (0.7);
        \draw[thick, green, style={decoration={markings, mark=at position 0.5 with {\arrow{>}}},postaction={decorate}}] (3,-0.45) circle (0.7);
        \draw[thick, magenta, style={decoration={markings, mark=at position 0.65 with {\arrow{>}}},postaction={decorate}}] (-3,-1.2) -- (3,-1.2);
        \draw[thick, green, style={decoration={markings, mark=at position 0. with {\arrow{<}}},postaction={decorate}}] (-3,1.35) circle (0.7);
        \draw[thick, green, style={decoration={markings, mark=at position 0.5 with {\arrow{>}}},postaction={decorate}}] (3,1.35) circle (0.7);
        \end{scope}
        \draw[->, thick] (-3.2,0) -- (3.2,0) node[right] {$\Re(z)$};
        \draw[->, thick] (0,-2.1) -- (0,2.1) node[above] {$\Im(z)$};
        \draw[dotted] (-3,0.9) -- (3,0.9);
        \draw[dashed] (-3,1.8) -- (3,1.8);
        \draw[dotted] (-3,-0.9) -- (3,-0.9);
        \draw[dashed] (-3,-1.8) -- (3,-1.8);
        \filldraw[black] (0,0.9) circle (1pt) node[below right] {$\frac{\pi}{2}i$};
        \filldraw[black] (0,1.8) circle (1pt) node[below right] {$\pi i$};
        \filldraw[black] (0,-0.9) circle (1pt) node[above right] {$-\frac{\pi}{2}i$};
        \filldraw[black] (0,-1.8) circle (1pt) node[above right] {$-\pi i$};
        \node at (2.82,0.45) {$-$};
        \node at (2.82,1.35) {$+$};
        \node at (-2.82,0.45) {$-$};
        \node at (-2.82,1.35) {$+$};
        \node at (2.82,-0.45) {$+$};
        \node at (2.82,-1.35) {$-$};
        \node at (-2.82,-0.45) {$+$};
        \node at (-2.82,-1.35) {$-$};
        \node at (-1,0.5) {$\gamma_{\Ai}$};
        \node at (-1,-1.5) {$\gamma'_{\Ai}$};
        \node at (-1.9,-0.5) {$\gamma_{\Bi}^{(1)}$};
        \node at (1.9,-0.5) {$\gamma_{\Bi}^{(2)}$};
        \node at (-1.9,1.35) {$\gamma'^{(1)}_{\Bi}$};
        \node at (1.9,1.35) {$\gamma'^{(2)}_{\Bi}$};
    \end{tikzpicture}
    \caption{Schematic illustration of the steepest ascent (orange) and steepest descent (blue) directions in the complex $p$-plane (left) and the complex $z$-plane (right) for the Salpeter equation. The left panel depicts only the positive sheet, while the negative sheet has the signs flipped. The magenta and green lines represent the cycles for the relativistic $\Ai$ and $\Bi$ functions, respectively. The branch cuts are indicated by the wavy lines.}
    \label{fig:sigma}
\end{figure}
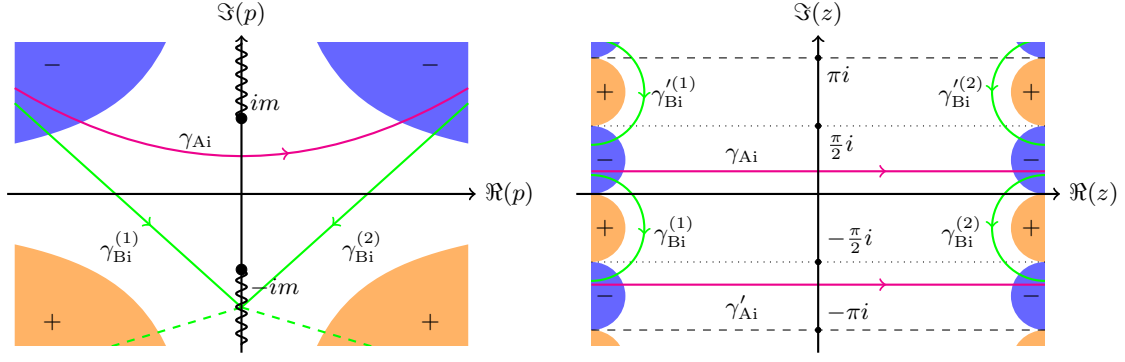

To facilitate the analysis, we introduce cylindrical coordinates on $\Sigma$ as follows:
\begin{equation}
    p = m \sinh z, \quad T = m \cosh z,\label{eq:z_s}
\end{equation}
where $\Re(z) \in \mathbb{R}$ and $\Im(z) \in [-\pi, \pi)$. The branch points on the original Riemann sheets are located at $z = \pm i\pi/2$. The standard branch cuts lie along $\Im(z) = \pm \pi/2$, and the positive Riemann sheet corresponds to $\Im(z) \in [-\pi/2, \pi/2)$. For an illustration of these coordinates, see Fig.~\ref{fig:sigma}.

We select an arbitrary base point for the integration at $ z = 0 $, corresponding to $ (p, T) = (0, m) $. The endpoint of the integration is naturally extended to $z \in \mathbb{C}$ such that the winding number of the cycle is given by $n = \lfloor \Im(z) / (2\pi) + 1/2 \rfloor$.
The exponent is computed as
\begin{align}
    S_S(z)=\frac{1}{g}\int_\gamma T\odif{p}=\frac{m^2}{g}\int_0^z\cosh^2z'\odif{z'}=\frac{m^2}{2g}\ab(z+\frac{1}{2}\sinh2z).
\end{align}

Let us examine the entire exponent part in the $z$ coordinates, $\mathcal I_S(x,z)=iS_S(z)+ixm\sinh z$.
We see that the integration over the non-trivial homology cycle gives monodromy of
\begin{equation}
    \mathcal I_S(x,z+2k\pi i)=\mathcal I_S(x,z)-k\frac{m^2\pi}{g},\label{eq:monodromy}
\end{equation}
for $k\in\mathbb Z$. In addition, for a half period, we have
\begin{equation}
    \mathcal I_S(x,z+\pi i)=\mathcal I_S(-x,z)-\frac{m^2\pi}{2g}.\label{eq:half-cycle}
\end{equation}
Although the Salpeter equation itself does not exhibit Klein tunneling, the suppression factor $-m^2\pi/(2g)$ is identical to that appearing in Klein tunneling; its imprint is nevertheless encoded in the structure of the exponent.

\subsection{General solutions}
We now turn to the classification of fast-decay cycles. As illustrated in Fig.~\ref{fig:sigma}, there are four steepest descent directions within a single period, and any cycle connecting two of these directions defines a solution. We adopt the basis for the cycles as $(\gamma_{\Ai} + 2\pi i n, \gamma'_{\Ai} + 2\pi i n, \gamma^{(1)}_{\Bi} + \gamma^{(2)}_{\Bi} + 2\pi i n, \gamma'^{(1)}_{\Bi} + \gamma'^{(2)}_{\Bi} + 2\pi i n)$ for $n \in \mathbb{Z}$, where the sum of cycles denotes their union and the addition of a complex number indicates a shift in the complex plane. Any fast-decay cycle can be decomposed as a linear combination of these basis cycles\footnote{There is an additional descent direction, $\Im(z) \to \infty$, though the decay of the exponent is relatively slow. This represents cycles wrapping around the cylinder infinitely many times, and the corresponding solutions turn out to be dependent on others.}.

Due to Eq.~\eqref{eq:monodromy}, we have
\begin{equation}
    \int_{\gamma+2\pi i n}\frac{\odif{z}}{2\pi}\odv{p}{z}e^{\mathcal I_S(x,z)}=e^{-n\frac{m^2\pi}{g}}\int_{\gamma}\frac{\odif{z}}{2\pi}\odv{p}{z}e^{\mathcal I_S(x,z)},
\end{equation}
where $\gamma$ is an arbitrary cycle and $\odv{p}{z} = m \cosh z$. Consequently, despite the infinite number of independent fast-decay cycles, the space of solutions is four-dimensional.
In addition, Eq.~\eqref{eq:half-cycle} gives
\begin{equation}
    \int_{\gamma+\pi i}\frac{\odif{z}}{2\pi}\odv{p}{z}e^{\mathcal I_S(x,z)}=-e^{-\frac{m^2\pi}{2g}}\int_{\gamma}\frac{\odif{z}}{2\pi}\odv{p}{z}e^{\mathcal I_S(-x,z)},
\end{equation}
which relates two solutions by the parity transformation.
Therefore, the general solution is given by four independent functions,
\begin{equation}\label{eq:general_sol}
    f(x)=c_{\Ai}\int_{\gamma_{\Ai}}\frac{\odif{z}}{2\pi}\odv{p}{z}e^{\mathcal I_S(x,z)}+c_{\Bi}\int_{\gamma_{\Bi}}\frac{\odif{z}}{2\pi}\odv{p}{z}e^{\mathcal I_S(x,z)}+c'_{\Ai}\int_{\gamma_{\Ai}}\frac{\odif{z}}{2\pi}\odv{p}{z}e^{\mathcal I_S(-x,z)}+c'_{\Bi}\int_{\gamma_{\Bi}}\frac{\odif{z}}{2\pi}\odv{p}{z}e^{\mathcal I_S(-x,z)}.
\end{equation}
The first and second solutions correspond to the classical $\Ai$ and $\Bi$ functions in the non-relativistic limit, respectively, while the last two correspond to their negative energy counterparts.
In Appendix \ref{apx:relativistic-airy}, we present a detailed analysis of these functions, demonstrating that their asymptotic behaviors differ and hence they are linearly independent.
The successful reconstruction of the complete solution space anticipated from the non-relativistic limit indicates the applicability of the Riemann-Hilbert correspondence to this system.
A rigorous proof of the correspondence would however require demonstrating that no analytic solutions exist that do not possess a non-relativistic limit, which is beyond the scope of this paper.

\subsection{Stokes lines and Lefschetz thimbles}
For each $x$, a natural basis of cycles is provided by the Lefschetz thimbles, namely the steepest-descent cycles attached to saddle points or branch points.
In this subsection, we analyze the structure of the Lefschetz thimbles and identify the associated Stokes lines, which delineate the regions where the thimble configuration undergoes qualitative changes.

First, note that the imaginary parts of the saddle points and branch points are degenerate across periods, which would ordinarily obstruct a straightforward application of Picard-Lefschetz theory. In the present analysis, however, this degeneracy does not affect the result, since the Stokes lines can be avoided for generic values of $x$.
This property is unique to the Salpeter equation, in contrast to the Dirac and Klein-Gordon equations.

The exponent $\mathcal{I}_S(x,z)$ possesses four critical points on the Riemann surface $\Sigma$:
\begin{equation}\label{eq:saddle_z}
    z_{S_\pm^{(n)}}=\pm\arccosh\ab(-\frac{gx}{m})+2n\pi i,\quad z_{B_\pm^{(n)}}=\pm\frac{\pi i}{2}+2n\pi i.
\end{equation}
The first two represent saddle points of the exponent in the original $p$ variable, $\mathcal{I}_S(x,p)$. They correspond to the classical momentum $p=\pm\sqrt{g^2x^2-m^2}$, but the branch-cut structure differs: $\arccosh(z)$ has a branch cut along $(-\infty,1)$ with $\Re(\arccosh(z))\geq0$ and $|\Im(\arccosh(z))|\leq\pi$ for $z\in\mathbb C$.
Notably, these saddle points accurately characterize the classical dynamics, despite the non-local nature of the Salpeter equation.
The latter two are branch points of the square-root operator, $p=\pm im$. Although they are not saddle points of the original exponent due to the divergence of $\odif{z}/\odif{p}$ at these locations, they effectively function as saddle points in the $z$-plane with a vanishing prefactor $\odif{p}/\odif{z}=m\cosh z=0$.
Although the Lefschetz thimble decomposition is unaffected by zero-points of the prefactor, the saddle-point expansion around these branch points requires the inclusion of higher-order corrections; see Appendix \ref{apx:relativistic-airy} for details.

\begin{figure}[t]
    \centering
    \includegraphics[width=0.4\linewidth]{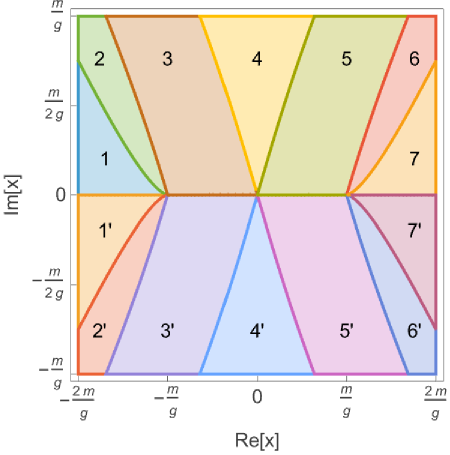}
    \caption{The Stokes lines for the exponent $\mathcal{I}_S$ under the conditions $m > 0$ and $g > 0$. The Lefschetz thimble structure remains invariant with respect to the specific real values of $m$ and $g$, and is depicted in Figs.~\ref{fig:thimbles14} and \ref{fig:thimbles57}. The relevant Stokes lines vary depending on the cycle; those for the relativistic Airy functions are given in Appendix \ref{apx:relativistic-airy}.}
    \label{fig:stokes}
\end{figure}

\begin{figure}[t]
    \centering
    \includegraphics[width=0.24\linewidth]{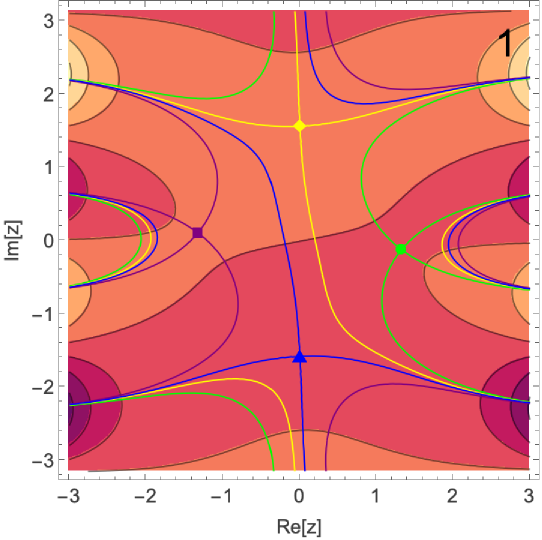}
    \includegraphics[width=0.24\linewidth]{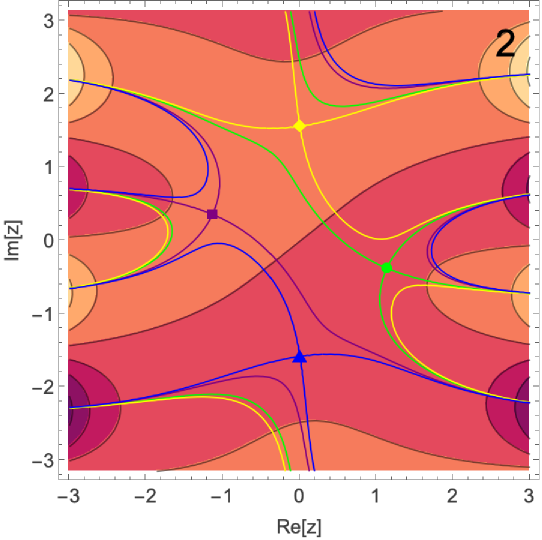}
    \includegraphics[width=0.24\linewidth]{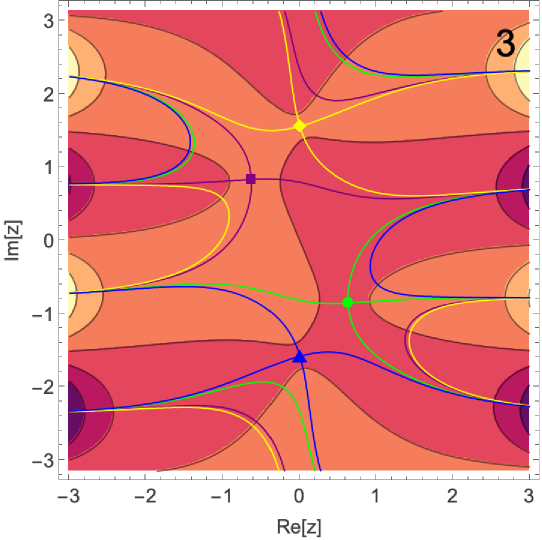}
    \includegraphics[width=0.24\linewidth]{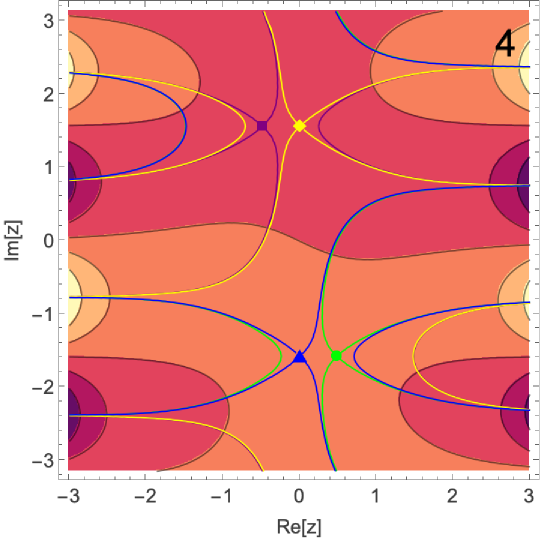}
    \includegraphics[width=0.24\linewidth]{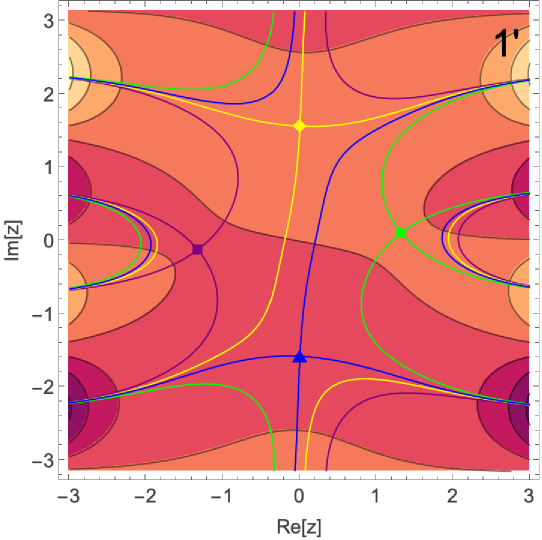}
    \includegraphics[width=0.24\linewidth]{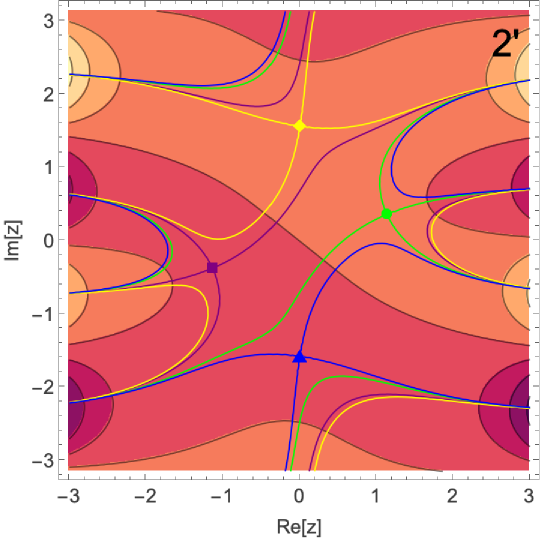}
    \includegraphics[width=0.24\linewidth]{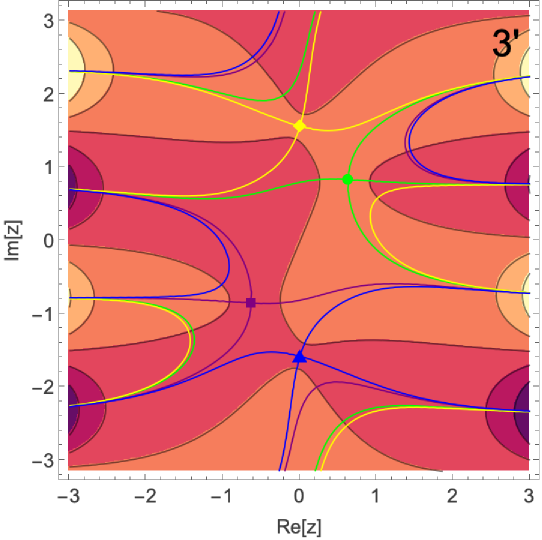}
    \includegraphics[width=0.24\linewidth]{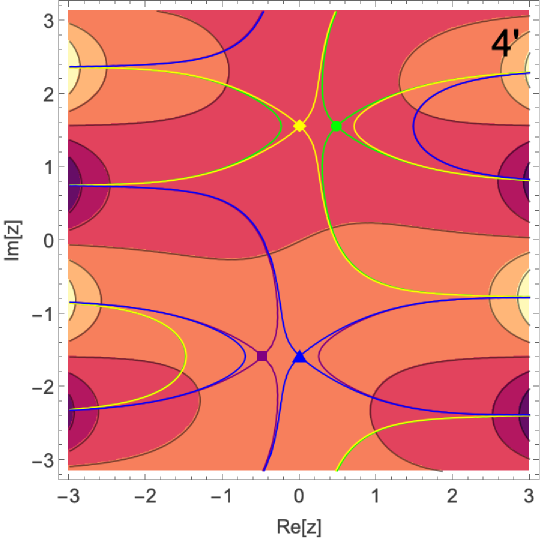}
    \caption{The structure of the Lefschetz thimbles for $\mathcal{I}_S$ in regions $1, 2, 3, 4, 1', 2', 3', 4'$. The parameters are set to $m = 5$ and $g = 0.1$, though the structure remains independent of these values provided they are real. Saddle points are marked by green circles ($S_+^{(0)}$) and purple squares ($S_-^{(0)}$), while branch points are indicated by yellow diamonds ($B_+^{(0)}$) and blue triangles ($B_-^{(0)}$). The contours represent lines of constant imaginary part of $\mathcal{I}_S$, with those connecting a saddle or branch point to a darker region denoting the Lefschetz thimbles. Shaded contours illustrate the real part of $\mathcal{I}_S$, where lighter regions correspond to large positive values and darker regions to large negative values. The figure repeats periodically in the imaginary direction with varying real parts of $\mathcal{I}_S$.}   \label{fig:thimbles14}
\end{figure}

\begin{figure}[t]
    \centering
    \includegraphics[width=0.24\linewidth]{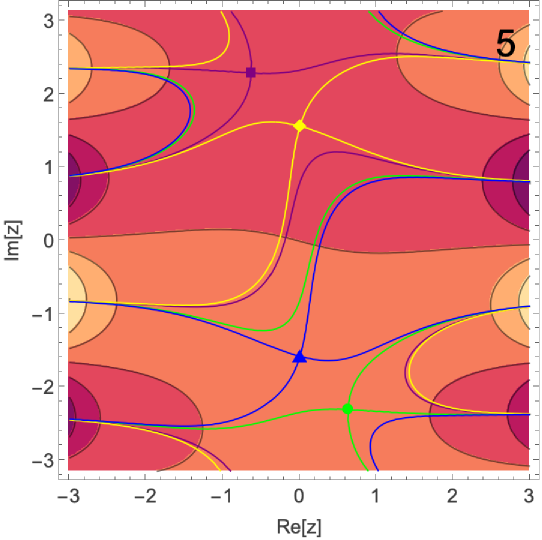}
    \includegraphics[width=0.24\linewidth]{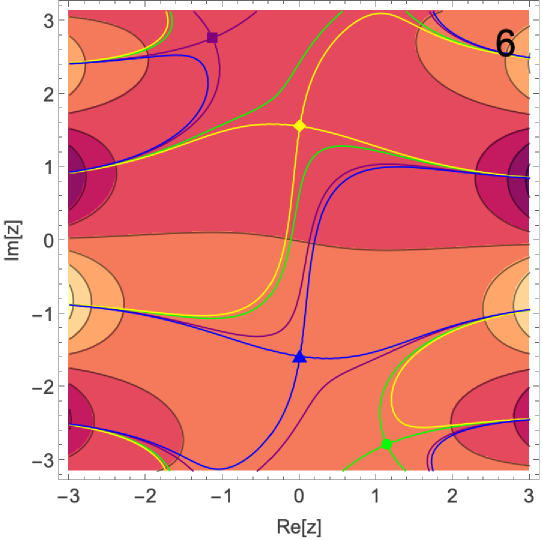}
    \includegraphics[width=0.24\linewidth]{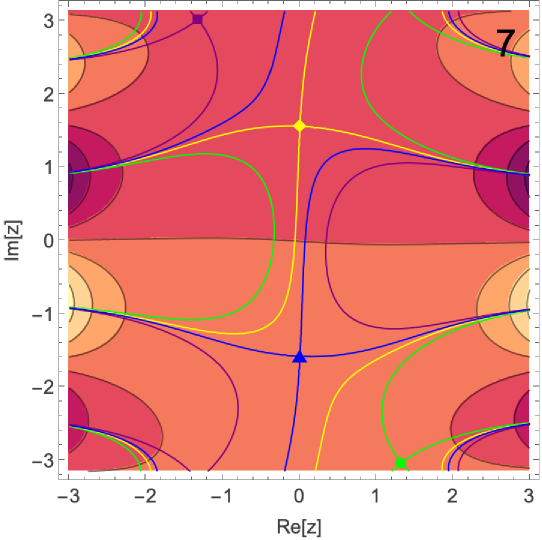}\\
    \includegraphics[width=0.24\linewidth]{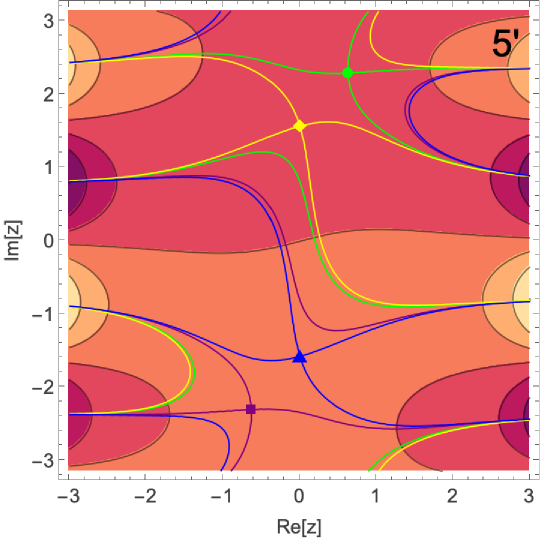}
    \includegraphics[width=0.24\linewidth]{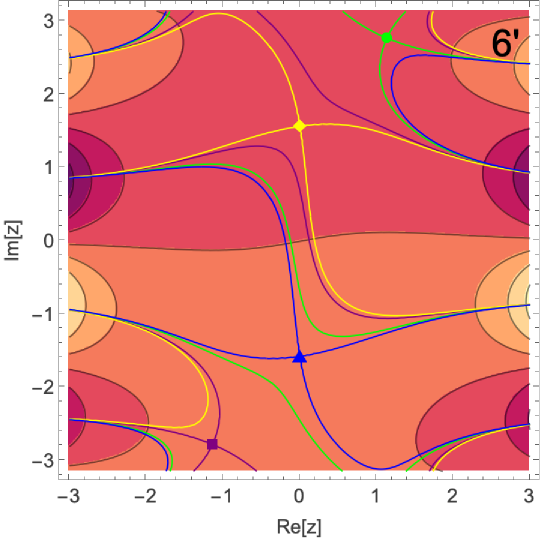}
    \includegraphics[width=0.24\linewidth]{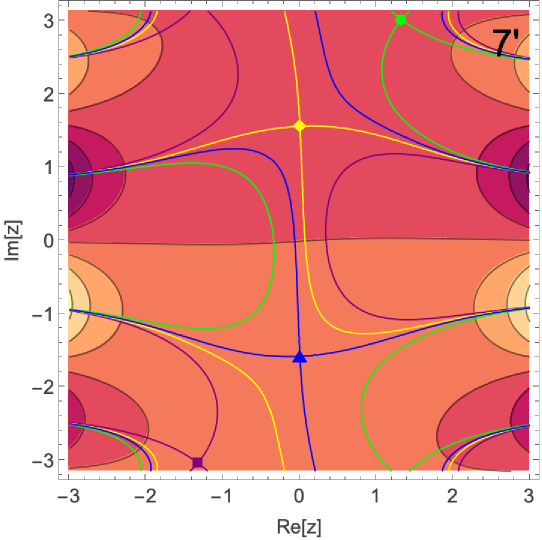}
    \caption{The same plots as in Fig.~\ref{fig:thimbles14}, but in regions $5,6,7,5',6',7'$.}
    \label{fig:thimbles57}
\end{figure}

In Fig.~\ref{fig:stokes}, we display all possible Stokes lines, where the imaginary parts of two saddle values coincide and the Lefschetz thimble structure changes.
For each region, the corresponding saddles, branch points, and Lefschetz thimbles are shown in Figs.~\ref{fig:thimbles14} and \ref{fig:thimbles57}.
In these figures, Lefschetz thimbles are the contours of $\Im(\mathcal I_S)$ connecting a saddle or branch point to regions of steepest descent; the remaining contours are not relevant.
The Lefschetz thimble structure is independent of the values of $g$ and $m$, provided that they are real.
For complex parameter values, the thimble topology undergoes structural transitions; the consequences of these modifications are explored in Sec.~\ref{sec:schwinger}.

\section{Lefschetz thimble analysis on Dirac and Klein-Gordon equations}\label{sec:dirac}
To understand the behavior of the Salpeter equation, we conduct a parallel Lefschetz thimble analysis for the Dirac and Klein-Gordon equations, which are known to exhibit Klein tunneling.
\subsection{Kernel for the Dirac equation}
We consider the time-independent one-dimensional Dirac equation with a linear potential,
\begin{equation}
    \ab[-i\alpha\partial+\beta m+gx]\psi=0,
\end{equation}
where\footnote{We employ the chiral representation where $\gamma^0 = \sigma_1$ and $\gamma^1 = -i\sigma_2$, with $\sigma_i$ denoting the Pauli matrices. The charge conjugation is defined as $\psi \to C\bar\psi^T$ with $C = -i\sigma_2$.}
\begin{equation}
    \alpha=
    \begin{pmatrix}
        1&0\\
        0&-1
    \end{pmatrix},\quad
    \beta=
    \begin{pmatrix}
        0&1\\
        1&0
    \end{pmatrix}.
\end{equation}
Here, the energy eigenvalue has been set to zero through a shift of $x$. The solutions to this equation are well-established in the literature and can be expressed in terms of parabolic cylinder functions \cite{Gavrilov:2007hq,Gavrilov:2015zem,Akhmedov:2020dgc}. In the following analysis, we reconstruct these solutions using Lefschetz thimbles to facilitate a direct comparison with the Salpeter equation.

We first perform the gauge transformation
\begin{equation}
    \psi\to \tilde \psi=e^{-\frac{ig}{2}x^2}\psi,~\partial\to D=\partial+igx,
\end{equation}
which yields
\begin{equation}
    \begin{pmatrix}
        -i\partial+2gx&m\\
        m&i\partial
    \end{pmatrix}\tilde \psi=0.
\end{equation}
We solve this by employing the Ansatz
\begin{equation}
    \tilde \psi_\gamma(x)=\int_\gamma\frac{\odif{k}}{2\pi}e^{ikx+iS_\psi(k)}
    \begin{pmatrix}
        1\\
        \frac{m}{k}
    \end{pmatrix}.
\end{equation}
Substituting this into the equation yields
\begin{equation}
    \odv{S_\psi}{k}=\frac{m^2+k^2}{2gk},
\end{equation}
which admits the solution
\begin{equation}
    S_\psi(k)=\frac{k^2}{4g}+\frac{m^2}{2g}\ln \frac{k}{m}.
\end{equation}
Returning to the original gauge, the solution corresponding to a fast-decay cycle $\gamma$ is expressed as\footnote{
One can also solve the equation with a different gauge transformation $\tilde\psi=e^{\frac{ig}{2}x^2}\psi$. This gives another representation of the solution:
\begin{equation}
    \psi(x)=\int_\gamma\frac{\odif{p}}{2\pi}e^{\mathcal I_{DKG}(x,p)|_{g\to-g}}
    \begin{pmatrix}
        -\frac{m}{p+gx}\\
        1
    \end{pmatrix}.
\end{equation}
}
\begin{equation}
    \psi_\gamma(x)=\int_\gamma\frac{\odif{p}}{2\pi}e^{\mathcal I_{DKG}(x,p)}
    \begin{pmatrix}
        1\\
        \frac{m}{p-gx}
    \end{pmatrix},
\end{equation}
where
\begin{equation}\label{eq:i_kg}
    \mathcal I_{DKG}(x,p)=i\frac{p^2}{4g}+i\frac{px}{2}-i\frac{gx^2}{4}+i\frac{m^2}{2g}\ln\frac{p-gx}{m}.
\end{equation}
Here, we used the canonical momentum $p=k+gx$, which agrees with the kinetic (or physical) momentum.
The singularity at $p=gx$ in the lower component poses no difficulty, as it only adds a constant to the coefficient of $\ln(p-gx)$ appearing in the exponent $\mathcal I_{DKG}(x,p)$.
The integrand is inherently multivalued, reflecting the monodromy associated with the logarithmic term in the exponent.

\subsection{Kernel for the Klein-Gordon equation}
Next, we consider the Klein-Gordon equation with a linear scalar potential,
\begin{equation}
    \ab[(\partial_t+igx)^2-\partial^2+m^2]\Phi(x,t)=0.
\end{equation}
Employing an energy eigenfunction $\Phi(x,t) = e^{-iEt}\phi_E(x)$ and eliminating the energy eigenvalue $E$ through a shift of $x$, we reduce to the equation
\begin{equation}
    \ab[-\partial^2+m^2-g^2x^2]\phi(x)=0.
\end{equation}
Following the same gauge transformation procedure as for the Dirac equation, $\phi \to \tilde{\phi} = e^{-\frac{ig}{2}x^2}\phi$, we obtain a solvable form. The general solution admits the integral representation
\begin{equation}
    \phi_\gamma(x)=\int_\gamma\frac{\odif{p}}{2\pi}\sqrt{\frac{m}{p-gx}}e^{\mathcal I_{DKG}(x,p)},
\end{equation}
where $\mathcal I_{DKG}(x,p)$ is given in Eq.~\eqref{eq:i_kg}.
The square root in the prefactor respects the monodromy of the logarithmic term in the exponent.
Importantly, the exponent is identical for both the Dirac and Klein-Gordon equations and the only difference is the prefactor.
\subsection{General solutions}
We now proceed to the classification of fast-decay cycles for the Dirac and Klein-Gordon equations. Given that the exponents are identical, the classification applies uniformly to both equations.
Note that one could incorporate the prefactor into the exponent, which would slightly displace the saddle points and Lefschetz thimbles, thereby altering the thimble decompositions near the Stokes lines. Nevertheless, the specific decomposition around the Stokes lines is a matter of convention and does not influence the final outcome.

\begin{figure}
    \centering
    \begin{tikzpicture}
        \begin{scope}
        \clip (-3,-2) -- (3,-2) -- (3,2) -- (-3,2) -- cycle;
        \fill[blue!60] plot[domain=-3:-0.1] (\x, {2/\x}) -- (-3,-3) -- cycle;
        \fill[orange!60] plot[domain=0.1:3] (\x, {-2/\x}) -- (3,-3) -- cycle;
        \fill[orange!60] plot[domain=-3:-0.1] (\x, {-2/\x}) -- (-3,3) -- cycle;
        \fill[blue!60] plot[domain=0.1:3] (\x, {2/\x}) -- (3,3) -- cycle;
        \draw[thick, green, style={decoration={markings, mark=at position 0.33 with {\arrow{>}}},postaction={decorate}}] (3,2) -- (-3,-3);
        \draw[thick, green, style={decoration={markings, mark=at position 0.5 with {\arrow{<}}},postaction={decorate}}] (3,2) -- (-2,0);
        \draw[thick, green, dashed] (-2,0) -- (-3,-2);
        \end{scope}
        \draw[->, thick] (-3.1,0) -- (3.1,0) node[right] {$\Re(p)$};
        \draw[->, thick] (0,-2.1) -- (0,2.1) node[above] {$\Im(p)$};
        \draw[thick, decorate, decoration = {snake,amplitude=2pt,segment length=5pt}] (-1,0) -- (-3,0);
        \filldraw[black] (-1,0) circle (2pt) node[below right] {$gx$};
        \node at (-2.5,1.7) {$+$};
        \node at (2.5,1.7) {$-$};
        \node at (2.5,-1.7) {$+$};
        \node at (-2.5,-1.7) {$- $};
        \node at (0.5,-0.5) {$\gamma_R$};
        \node at (-0.5,1) {$\gamma'_R$};
    \end{tikzpicture}
    \quad
    \begin{tikzpicture}
        \begin{scope}
        \clip (-3,-2) -- (3,-2) -- (3,2) -- (-3,2) -- cycle;
        \fill[orange!60] (3,1.35) circle (0.45);
        \fill[blue!60] (3,0.45) circle (0.45);
        \fill[blue!60] (3,-1.35) circle (0.45);
        \fill[orange!60] (3,-0.45) circle (0.45);
        \fill[blue!60] (3,2.25) circle (0.45);
        \fill[orange!60] (3,-2.25) circle (0.45);
        \draw[thick, green, style={decoration={markings, mark=at position 0.5 with {\arrow{>}}},postaction={decorate}}] (3,-0.45) circle (0.7);
        \draw[thick, green, style={decoration={markings, mark=at position 0.5 with {\arrow{>}}},postaction={decorate}}] (3,1.35) circle (0.7);
        \end{scope}
        \draw[->, thick] (-3.2,0) -- (3.2,0) node[right] {$\Re(z)$};
        \draw[->, thick] (0,-2.1) -- (0,2.1) node[above] {$\Im(z)$};
        \draw[dotted] (-3,0.9) -- (3,0.9);
        \draw[dashed] (-3,1.8) -- (3,1.8);
        \draw[dotted] (-3,-0.9) -- (3,-0.9);
        \draw[dashed] (-3,-1.8) -- (3,-1.8);
        \filldraw[black] (0,0.9) circle (1pt) node[below right] {$\frac{\pi}{2}i$};
        \filldraw[black] (0,1.8) circle (1pt) node[below right] {$\pi i$};
        \filldraw[black] (0,-0.9) circle (1pt) node[above right] {$-\frac{\pi}{2}i$};
        \filldraw[black] (0,-1.8) circle (1pt) node[above right] {$-\pi i$};
        \node at (2.82,0.45) {$-$};
        \node at (2.82,1.35) {$+$};
        \node at (2.82,-0.45) {$+$};
        \node at (2.82,-1.35) {$-$};
        \node at (2,-0.5) {$\gamma_R$};
        \node at (2,1.35) {$\gamma'_R$};
    \end{tikzpicture}
    \caption{Schematic illustration of the steepest ascent (orange) and descent (blue) directions in the complex $p$-plane (left) and the complex $z$-plane (right) for the Dirac and Klein-Gordon equations. The topology of the steepest ascent and descent regions is uniform across all Riemann sheets. The green lines show two independent cycles. The branch cut is indicated by the wavy line.}
    \label{fig:sigmaKG}
\end{figure}
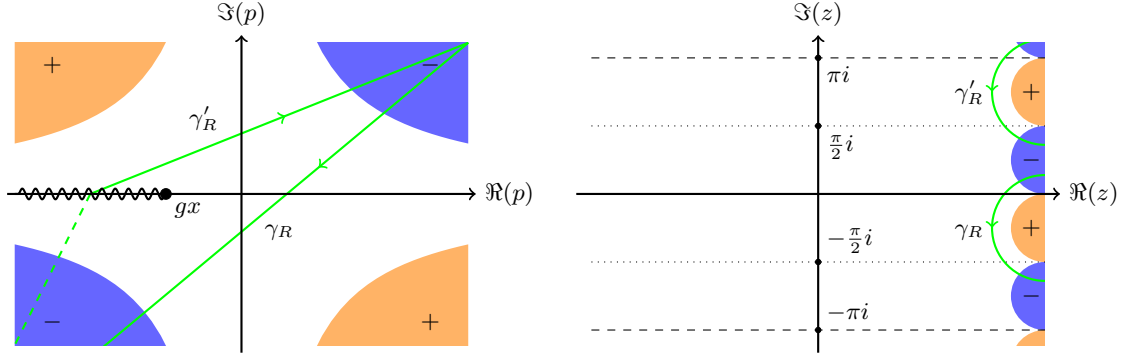

For the analysis of the Lefschetz thimbles, we employ the coordinate transformation defined by
\begin{equation}
    p = m e^{z} + g x,\label{eq:z_dkg}
\end{equation}
which is illustrated in Fig.~\ref{fig:sigmaKG}.
Notably, the exponent $\mathcal{I}_{DKG}(x,z)$ exhibits the same monodromy properties as that of the Salpeter equation, as encapsulated in Eqs.~\eqref{eq:monodromy} and \eqref{eq:half-cycle}.

Since we have two steepest descent directions within $\Im (z) \in [-\pi, \pi)$, we adopt the basis for the fast-decay cycles as $(\gamma_R + 2\pi i n, \gamma'_R + 2\pi i n)$ for $n \in \mathbb{Z}$.
Then, Eq.~\eqref{eq:monodromy} implies that the space of solutions is two-dimensional, in agreement with the second-order character of the underlying differential equations.
Using Eq.~\eqref{eq:half-cycle}, we obtain
\begin{equation}
    \psi(x) = c_\psi \int_{\gamma_R} \frac{\odif{z}}{2\pi} \odv{p}{z} \begin{pmatrix} 1 \\ e^{-z} \end{pmatrix} e^{\mathcal{I}_{DKG}(x,z)} + c'_\psi \int_{\gamma_R} \frac{\odif{z}}{2\pi} \odv{p}{z} \begin{pmatrix} 1 \\ e^{-z} \end{pmatrix} e^{\mathcal{I}_{DKG}(-x,z)},
\end{equation}
for the Dirac equation, and
\begin{equation}
    \phi(x) = c_\phi \int_{\gamma_R} \frac{\odif{z}}{2\pi} \odv{p}{z} e^{-\frac{z}{2}} e^{\mathcal{I}_{DKG}(x,z)} + c'_\phi \int_{\gamma_R} \frac{\odif{z}}{2\pi} \odv{p}{z} e^{-\frac{z}{2}} e^{\mathcal{I}_{DKG}(-x,z)},
\end{equation}
for the Klein-Gordon equation. Here, $\odv{p}{z} = m e^{z}$.

\subsection{Stokes lines and Lefschetz thimbles}
Within a single period, there exist precisely two saddle points that coincide with $z_{S_\pm^{(n)}}$ in Eq.~\eqref{eq:saddle_z}. The corresponding momentum values are also identical, despite the distinct coordinate transformations defined in Eqs.~\eqref{eq:z_s} and \eqref{eq:z_dkg}.
Notably, in the $z$ coordinates, the exponent for the Salpeter equation $\mathcal I_S$ can be obtained by integrating
\begin{equation}
    \pdv{\mathcal I_S}{z}=\frac{1+e^{-2z}}{2}\pdv{\mathcal I_{DKG}}{z}.
\end{equation}
Here, the factor $1+e^{-2z}$ introduces the branch points $z_{B_\pm^{(n)}}$. Consequently, $\mathcal I_S$ may be interpreted as a minimal deformation of $\mathcal I_{DKG}$ that incorporates these branch points.

\begin{figure}[t]
    \centering
    \includegraphics[width=0.4\linewidth]{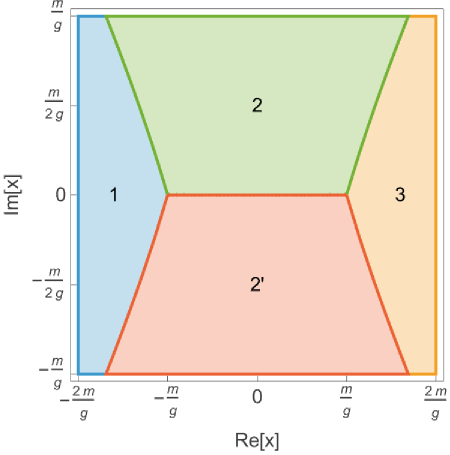}
    \caption{The Stokes lines for $\mathcal I_{DKG}$ in the limit where the imaginary parts of $m$ and $g$ go to zero. The structure of the Stokes lines is independent of the specific real values of $m$ and $g$, and the Lefschetz thimbles for each region are depicted in Fig.~\ref{fig:thimblesKG}.}
    \label{fig:stokesKG}
\end{figure}

\begin{figure}[t]
    \centering
    \includegraphics[width=0.24\linewidth]{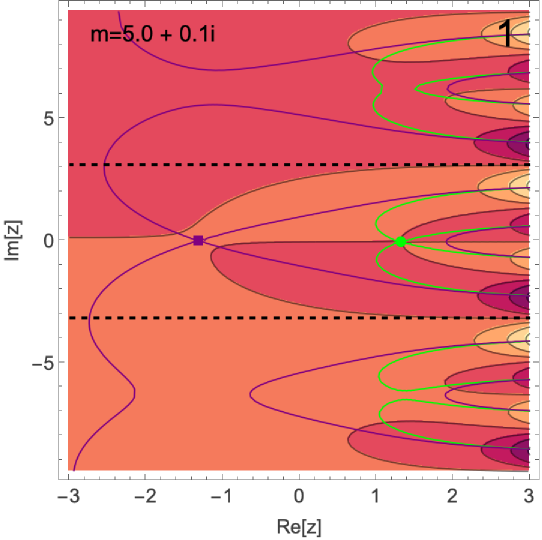}
    \includegraphics[width=0.24\linewidth]{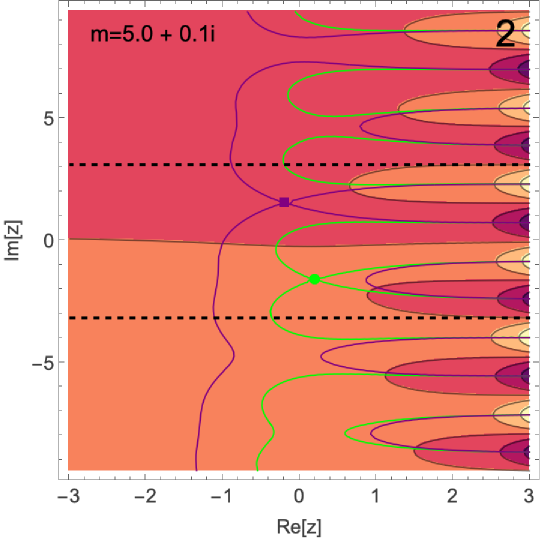}
    \includegraphics[width=0.24\linewidth]{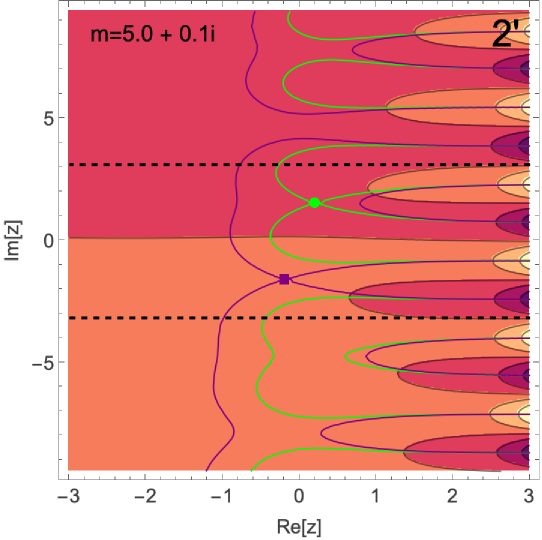}
    \includegraphics[width=0.24\linewidth]{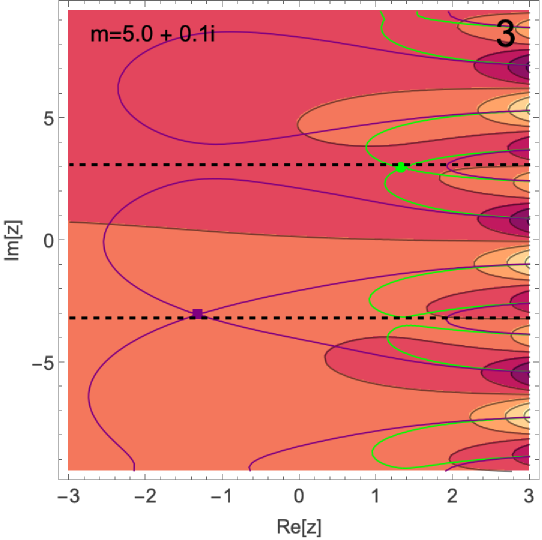}
    \includegraphics[width=0.24\linewidth]{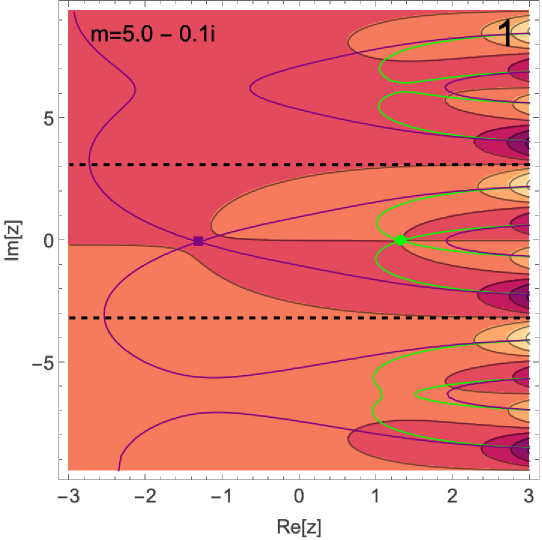}
    \includegraphics[width=0.24\linewidth]{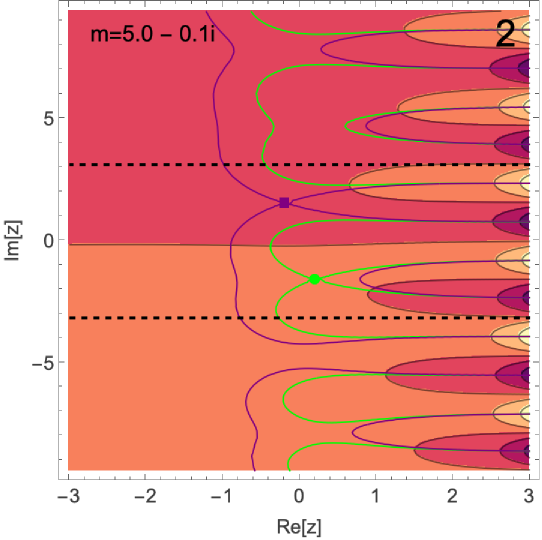}
    \includegraphics[width=0.24\linewidth]{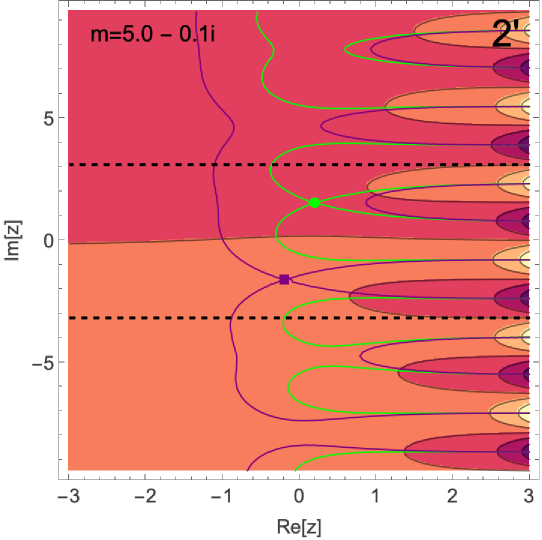}
    \includegraphics[width=0.24\linewidth]{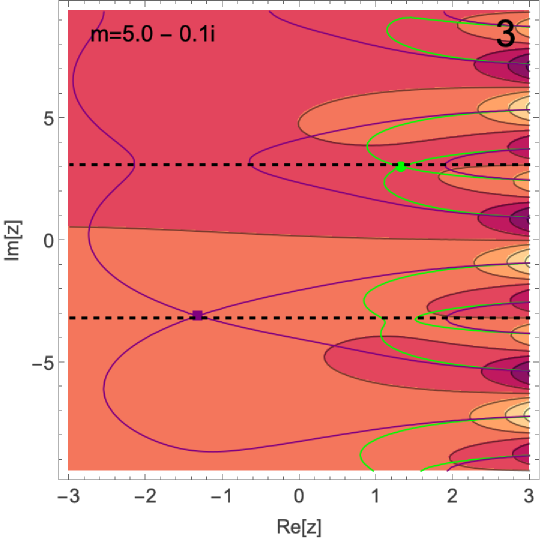}
    \caption{The structure of the Lefschetz thimbles for $\mathcal I_{KG}$. While the specific values $m=5\pm 0.1i$ and $g=0.1$ are used for illustration, the overall thimble structure remains invariant if the imaginary parts of these parameters are small enough. The saddle points are indicated by green circles ($S_+^{(0)}$) and purple squares ($S_-^{(0)}$). Three periods are shown, delineated by dashed lines. The contour and shading conventions follow those established in Fig.~\ref{fig:thimbles14}.}
    \label{fig:thimblesKG}
\end{figure}

The absence of the branch points significantly alters the behavior of the solutions.
We illustrate the Stokes lines for $\mathcal I_{DKG}$ in Fig.~\ref{fig:stokesKG}.
Since the saddle points $z_{S_\pm^{(n)}}$ are identical, the Stokes lines illustrated in Fig.~\ref{fig:stokesKG} constitute a subset of those presented in Fig.~\ref{fig:stokes}. In Fig.~\ref{fig:thimblesKG}, we show the structure of the Lefschetz thimbles for each region.
Unlike for the Salpeter equation, the thimbles connect the saddles of $S_-^{(n)}$ in different  periods when $m$ and $g$ are real, implying that these parameters are on the Stokes line. Accordingly, small imaginary parts must be introduced to these parameters. We set $g=0.1$ and illustrate two alternatives for avoiding the Stokes line with $m=5\pm0.1i$. If $\Im(m)>0$, the Lefschetz thimble associated with $S_-^{(n)}$ connects two steepest descent directions. If $\Im(m)<0$, it connects a steepest descent direction with $\Im(z)=\infty$.
In contrast, the Lefschetz thimbles associated with $S_+^{(n)}$ consistently connect two steepest descent directions. This connectivity is analogous to that of the Salpeter equation, except in intermediate regions 2', 3', 5, and 6 in Fig.~\ref{fig:stokes}.

\section{Klein paradox and Schwinger effect}\label{sec:schwinger}
In this section, we investigate the Klein paradox and the Schwinger effect utilizing the results in the preceding sections.
We also introduce a topological blocking mechanism that accounts for the absence of these phenomena in the Salpeter equation.

\subsection{Dirac equation and Klein-Gordon equation}
The Klein paradox has historically been subject to conceptual ambiguity, necessitating a clear distinction between the group velocity and the conserved charge current. While asymptotic in- and out-states are characterized by their group velocities, the transmission and reflection coefficients are fundamentally determined by the flux of the charge current.
We recall that the charge densities for the Dirac and Klein-Gordon equations are defined respectively as
\begin{equation}
    \rho_\psi = \psi^\dagger \psi, \quad \rho_\phi = i(\phi^* \partial_t \phi - \phi \partial_t \phi^*) - 2gx |\phi|^2,
\end{equation}
with the corresponding currents given by
\begin{equation}
    j_\psi = \psi^\dagger \alpha \psi, \quad j_\phi = -i(\phi^* \partial \phi - \phi \partial \phi^*).
\end{equation}
Notably, since the Klein-Gordon equation is second-order in time, $|\phi|^2$ does not represent a probability density; rather, the conserved quantity is the spatial integral of the charge density $\rho_\phi$.
Furthermore, $\rho_\phi$ is not positive-definite and changes sign depending on the sign of $x$. This implies that the directions of the group velocity and the charge current are opposite in the region $x > 0$.

We consider tunneling from region 3 toward region 1, as the latter avoids the aforementioned complication in identifying outgoing waves. Then, the Klein tunneling is interpreted as an antiparticle incident from $x \to \infty$ that is transmitted through the potential barrier and emerges as a particle as $x \to -\infty$.
In the following analysis, we set $E=0$ without loss of generality, as a non-zero energy eigenvalue merely corresponds to a spatial translation.
Furthermore, we assume $\Im(m) > 0$, noting that we are interested in the limit $\Im(m)\to0$ and the final physical results remain invariant regardless of this choice. 

For $x\to-\infty$, the purely outgoing wave solution is associated with the cycle passing exclusively through the saddle point $S^{(0)}_-$, namely $\gamma'_R + \gamma_R$.
Note that the group velocity is determined by evaluating the energy dependence of the exponent under the shift $x \to x - E/g$.
The manifestation of the Klein paradox can be analyzed by examining the solution in the asymptotic limits $x \to \pm \infty$. For $x \to -\infty$, the saddle-point approximation yields
\begin{equation}
    \psi_{\gamma'_R+\gamma_R}(x)\simeq M_\psi e^{i\ab(\frac{g}{2}x^2-\frac{m^2}{2g}-\frac{\pi}{4})}\sqrt{\frac{g}{\pi}}\ab(-\frac{m}{2gx})^{i\frac{m^2}{2g}}\begin{pmatrix}
        -\frac{m}{2gx}\\
        1
    \end{pmatrix},\label{eq:dirac_left}
\end{equation}
for the Dirac equation, and
\begin{equation}
    \phi_{\gamma'_R+\gamma_R}(x)\simeq M_\phi e^{i\ab(\frac{g}{2}x^2-\frac{m^2}{2g}-\frac{\pi}{4})}\sqrt{\frac{g}{\pi}}\ab(-\frac{m}{2gx})^{i\frac{m^2}{2g}+\frac12},\label{eq:kg_left}
\end{equation}
for the Klein-Gordon equation.
Here, $M_\psi$ and $M_\phi$ are factors that compensate for the breakdown of the Gaussian approximation, to be determined subsequently.
This is caused by the interference among the saddle points $S^{(n)}_-$, which are connected by the Lefschetz thimbles in the limit $\Im(m) \to 0$.
In the $m^2/g \to \infty$ limit, these saddle points exhibit localized Gaussian profiles, and the factors $M_\psi$ and $M_\phi$ approach unity. 

For $x\to\infty$, the saddle-point expansion yields
\begin{align}
    \psi_{\gamma'_R+\gamma_R}(x)&\simeq M_\psi e^{\frac{m^2\pi}{2g}}e^{i\ab(\frac{g}{2}x^2-\frac{m^2}{2g}-\frac{\pi}{4})}\sqrt{\frac{g}{\pi}}\ab(\frac{m}{2gx})^{i\frac{m^2}{2g}}\begin{pmatrix}
        -\frac{m}{2gx}\\
        1
    \end{pmatrix}\nonumber\\
    &\hspace{3ex}-e^{\frac{m^2\pi}{2g}}\ab(1-e^{-\frac{m^2\pi}{g}})e^{-i\ab(\frac{g}{2}x^2-\frac{\pi}{4})}\sqrt{\frac{g}{\pi}}\ab(\frac{m}{2gx})^{-i\frac{m^2}{2g}}
    \begin{pmatrix}
        1\\
        -\frac{m}{2gx}
    \end{pmatrix},\label{eq:dirac_right}
\end{align}
for the Dirac equation, and
\begin{align}
    \phi_{\gamma'_R+\gamma_R}(x)&\simeq -M_\phi e^{\frac{m^2\pi}{2g}}e^{i\ab(\frac{g}{2}x^2-\frac{m^2}{2g}+\frac{\pi}{4})}\sqrt{\frac{g}{\pi}}\ab(\frac{m}{2gx})^{i\frac{m^2}{2g}+\frac12}+e^{\frac{m^2\pi}{2g}}\ab(1+e^{-\frac{m^2\pi}{g}})e^{-i\ab(\frac{g}{2}x^2+\frac{\pi}{4})}\sqrt{\frac{g}{\pi}}\ab(\frac{m}{2gx})^{-i\frac{m^2}{2g}+\frac12},\label{eq:kg_right}
\end{align}
for the Klein-Gordon equation.
Note that, by virtue of Eq.~\eqref{eq:half-cycle}, the coefficients $M_\psi$ and $M_\phi$ coincide in both asymptotic limits $x \to \pm \infty$.
Unlike $S_-^{(n)}$, the saddle-point approximation for $S_+^{(n)}$ becomes increasingly accurate in the limit $x \to \pm\infty$, since the second derivative $\odv[order=2]{\mathcal{I}_{DKG}}{z}$ at the saddle point scales as $igx^2$.
The difference between the asymptotic behaviors of the Dirac and Klein-Gordon equations is attributable to the monodromy properties of the prefactors; the Dirac prefactor remains invariant under the transformation $z \to z - 2\pi i$, while the Klein-Gordon prefactor changes sign.

The magnitudes $|M_\psi|$ and $|M_\phi|$ are fixed by current conservation, $j_{\psi/\phi}(\infty)=j_{\psi/\phi}(-\infty)$, which yields
\begin{equation}
    |M_\psi|=\sqrt{1-e^{-\frac{m^2\pi}{g}}},\quad |M_\phi|=\sqrt{1+e^{-\frac{m^2\pi}{g}}}.
\end{equation}
The phases do not affect the final results.
In Appendix \ref{apx:breakdown}, we numerically verify the accuracy of these expressions.

The preceding analysis facilitates the determination of the transmission coefficients.
For Eqs.~\eqref{eq:dirac_right} and \eqref{eq:kg_right}, the first term is incoming, and the second term is outgoing, as is determined by the group velocity. Meanwhile, the direction of the current is flipped in the region $x>0$ for the Klein-Gordon equation.
Therefore, the transmission coefficients are given by
\begin{equation}
    T_\psi=e^{-\frac{m^2\pi}{g}},\quad T_\phi=-e^{-\frac{m^2\pi}{g}},
\end{equation}
and the reflection coefficients are given by
\begin{equation}
    R_\psi=1-e^{-\frac{m^2\pi}{g}},\quad R_\phi=1+e^{-\frac{m^2\pi}{g}}.
\end{equation}
For the Klein-Gordon equation, we note that $T_\phi<0$ and $R_\phi>1$, a consequence of the charge sign reversal.
Remarkably, the magnitudes of the transmission coefficients increase as the potential barrier becomes steeper, approaching $T_\psi\to1$ and $T_\phi\to-1$ as $g\to\infty$. This counterintuitive behavior constitutes the Klein paradox, whose resolution requires second quantization.

We next investigate the Schwinger effect, in which the Klein paradox is interpreted as the production of particle-antiparticle pairs. While various methodologies exist for deriving the production rate, we employ the derivation through the Bogoliubov transformation \cite{Hansen_1981}.
To construct the full $S$-matrix, we examine an additional solution corresponding to the cycle $\gamma_R$.
For $x \to -\infty$, the saddle-point approximation yields
\begin{equation}
    \psi_{\gamma_R}(x)\simeq -e^{-i\ab(\frac{g}{2}x^2-\frac{\pi}{4})}\sqrt{\frac{g}{\pi}}\ab(-\frac{m}{2gx})^{-i\frac{m^2}{2g}}\begin{pmatrix}
        1\\
        -\frac{m}{2gx}
    \end{pmatrix},\label{eq:dirac_left_g}
\end{equation}
for the Dirac equation, and
\begin{equation}
    \phi_{\gamma_R}(x)\simeq -e^{-i\ab(\frac{g}{2}x^2-\frac{\pi}{4})}\sqrt{\frac{g}{\pi}}\ab(-\frac{m}{2gx})^{-i\frac{m^2}{2g}+\frac12},\label{eq:kg_left_g}
\end{equation}
for the Klein-Gordon equation.
Meanwhile, for $x\to\infty$, the saddle-point expansion yields
\begin{align}
    \psi_{\gamma_R}(x)&\simeq M_\psi e^{\frac{m^2\pi}{2g}}e^{i\ab(\frac{g}{2}x^2-\frac{m^2}{2g}-\frac{\pi}{4})}\sqrt{\frac{g}{\pi}}\ab(\frac{m}{2gx})^{i\frac{m^2}{2g}}\begin{pmatrix}
        -\frac{m}{2gx}\\
        1
    \end{pmatrix}-e^{\frac{m^2\pi}{2g}}e^{-i\ab(\frac{g}{2}x^2-\frac{\pi}{4})}\sqrt{\frac{g}{\pi}}\ab(\frac{m}{2gx})^{-i\frac{m^2}{2g}}
    \begin{pmatrix}
        1\\
        -\frac{m}{2gx}
    \end{pmatrix},\label{eq:dirac_right_g}
\end{align}
for the Dirac equation, and
\begin{align}
    \phi_{\gamma_R}(x)&\simeq -M_\phi e^{\frac{m^2\pi}{2g}}e^{i\ab(\frac{g}{2}x^2-\frac{m^2}{2g}+\frac{\pi}{4})}\sqrt{\frac{g}{\pi}}\ab(\frac{m}{2gx})^{i\frac{m^2}{2g}+\frac12}+e^{\frac{m^2\pi}{2g}}e^{-i\ab(\frac{g}{2}x^2+\frac{\pi}{4})}\sqrt{\frac{g}{\pi}}\ab(\frac{m}{2gx})^{-i\frac{m^2}{2g}+\frac12},\label{eq:kg_right_g}
\end{align}
for the Klein-Gordon equation.
Since the solution space is two-dimensional, the asymptotic behavior of the two independent solutions determine the connection formulas and the $S$-matrix.

In Appendix~\ref{apx:Bogoliubov}, we define the incoming and outgoing solutions for particles and antiparticles, and construct the $S$-matrix relating them. This allows for the determination of the Bogoliubov coefficients $\alpha_{\psi/\phi,E}$ and $\beta_{\psi/\phi,E}$, which are defined via the relation
\begin{align}
    \hat{a}_{L,E}^{\psi/\phi,\rm out} &= \alpha_{\psi/\phi,E} \hat a_{L,E}^{\psi/\phi,\rm in} + \beta_{\psi/\phi,E} \hat a_{R,E}^{\psi/\phi,\rm in\dagger}.
\end{align}
Here, $\hat a_{L/R,E}^{\psi/\phi,\rm in/out}$ represent the annihilation operators for the in- and out-states in the asymptotic regions $x \to -\infty$ ($L$) and $x \to \infty$ ($R$). These operators are normalized such that the canonical (anti-)commutation relations for identical labels yield the delta function $\delta(E-E')$. As derived in Appendix~\ref{apx:Bogoliubov}, these coefficients are given by:
\begin{equation}
    \alpha_{\psi,E}=-ie^{-i\frac{m^2-2E^2}{2g}}M_\psi,\quad \beta_\psi=e^{-\frac{m^2\pi}{2g}},
\end{equation}
for the Dirac equation, and
\begin{equation}
    \alpha_{\phi,E}=ie^{-i\frac{m^2-2E^2}{2g}}M_\phi,\quad \beta_\phi=ie^{-\frac{m^2\pi}{2g}},
\end{equation}
for the Klein-Gordon equation.
They satisfy $|\alpha_{\psi,E}|^2+|\beta_{\psi,E}|^2=1$ and $|\alpha_{\phi,E}|^2-|\beta_{\phi,E}|^2=1$, consistent with unitary evolution and the (anti-)commutation relations.

We proceed to evaluate the pair-production rate.
The in-state vacuum is defined by $\hat a_{L/R,E}^{\psi/\phi,\rm in}\ket|0>_{\rm in}=0$.
For the energy window $0<E<gl-m$ with $l>m/g$, pair production occurs in the region $m/g<|x|<l$.
Using the identity $\delta(E=0) = \lim_{T \to \infty} T / (2\pi)$, the pair production rate per unit length becomes
\begin{equation}
    \Gamma_{\psi/\phi}=\lim_{l\to\infty}\frac{1}{2\pi l\delta(0)}\int_0^{gl-m}\odif{E}\,{}_{\rm in}\bra<0|\hat a_{L,E}^{\psi/\phi,{\rm out}\dagger} \hat a_{L,E}^{\psi/\phi,{\rm out}}\ket|0>_{\rm in}=\frac{g}{2\pi}e^{-\frac{m^2\pi}{g}}.
\end{equation}
We note that while the calculation explicitly evaluates the particle count, the result for antiparticles is identical due to the symmetric nature of the pair-production process.
This result agrees with established calculations \cite{PhysRev.82.664,Sauter:1931zz,Heisenberg:1936nmg,Weisskopf:406571,Hansen_1981} in the single-instanton approximation.
Note that the corresponding three-dimensional rates are obtained by including the appropriate transverse phase-space factors.

\subsection{Salpeter equation}
For the Salpeter equation, the charge density is given by $\rho_f=|f|^2$, and the corresponding charge current $j_f$ satisfies $\partial j_f=if^*\sqrt{-\partial^2+m^2}f-if\sqrt{-\partial^2+m^2}f^*$; see \cite{Kowalski:2011zi} for further details.
In particular, for $f(x)=f_\gamma(x)$, we have
\begin{equation}
    j_f(x)=\int_{\gamma}\frac{\odif{z}}{2\pi}\int_{\bar\gamma}\frac{\odif{z'}}{2\pi}m^2\cosh(z')\cosh(z)\tanh\ab(\frac{z+z'}{2})e^{\mathcal I_S(x,z)-\mathcal I_S(x,z')},
\end{equation}
where $\bar\gamma$ is the cycle in which each point is the complex conjugate of the corresponding point of $\gamma$.
Since the exponent is separable with respect to the integration variables, we can apply the saddle point approximation to each variable. Note that the upward and downward flows are interchanged for the $z'$ integral.
When $\gamma$ is taken to be Lefschetz thimbles associated with $S_\pm^{(n)}$, we find that $j_f(x)\propto \pm g/(2\pi)$, while those associated with $B_\pm^{(n)}$ yield $j_f(x)\simeq0$. Notice that the direction of the group velocity is consistent with that of the charge current.

Let us examine the asymptotic behavior of the exponent at the saddle and branch points. In the limit $x \to \pm\infty$, the behavior at the saddle points is asymptotically given by
\begin{equation}
    \mathcal I_S(x,S_\pm^{(n)})= \mp i\frac{gx^2}{2}\pm i\frac{m^2}{2g}\ln (-x)+\order(1),\label{eq:saddle_S}
\end{equation}
which matches the behavior of $\mathcal I_{DKG}(x,S_\pm^{(n)})$.
Accordingly, the contributions from these saddles are normalizable.
In contrast, the behavior at the branch points is given by
\begin{equation}
    \mathcal I_S(x,B_\pm^{(n)})= \mp mx+\order(1).\label{eq:saddle_B}
\end{equation}
Consequently, in the limit $x\to\mp\infty$, the contributions from $B_\pm^{(n)}$ must vanish to ensure the solution remains bounded.

We now construct a solution satisfying purely outgoing boundary conditions as $x \to -\infty$ to investigate Klein tunneling. The above analysis restricts the solution to contributions from $S_-^{(n)}$ and possibly $B_-^{(n)}$, corresponding to the cycles $\gamma_{\Bi}^{(1)}$ and $\gamma'_{\Ai}$. However, any such cycle necessarily includes a divergent contribution from $B_-^{(n)}$ as $x\to\infty$, so no normalizable wave function exists that is composed solely of outgoing waves at $x\to-\infty$.
This conclusion is consistent with the asymptotic behavior of the relativistic Airy functions in Appendix \ref{apx:relativistic-airy}, where the only normalizable solutions are the relativistic $\Ai$ functions.

Let us compare the Lefschetz thimble structures between the Salpeter equation (Figs.~\ref{fig:thimbles14} and \ref{fig:thimbles57}) and the Dirac and Klein-Gordon equations (Fig.~\ref{fig:thimblesKG}). For $x\to\pm\infty$, the thimble from $S^{(n)}_+$ connects the same steepest descent directions, whereas the thimble from $S^{(n)}_-$ exhibits distinctly different behavior.
In the Salpeter case, this thimble connects steepest descent directions that are absent in the Dirac and Klein-Gordon case. Conversely, for the Dirac and Klein-Gordon equations, the thimble extends across adjacent periods.
Such inter-period connectivity is a prerequisite for Klein tunneling. In the Salpeter equation, the corresponding cycle is instead formed from the thimbles attached to the branch points $B_\pm^{(n)}$.
Specifically, in the absence of divergent branch points, {\it i.e.} $B_+^{(n)}$ for $x\to-\infty$ and $B_-^{(n)}$ for $x\to\infty$, the Lefschetz thimbles are unable to bridge steepest descent directions between distinct periods. Consequently, the admissible cycles are confined to the strips $-\pi+2\pi n \leq \Im(z) < \pi+2\pi n$ for $x\to-\infty$ and $2\pi n \leq \Im(z) < 2\pi+2\pi n$ for $x\to\infty$. To ensure the boundedness of the solution, the cycle must lie within the intersection of these regions, effectively restricting the solutions to the relativistic $\Ai$ functions.
This observation provides a geometric interpretation of the blocking mechanism for the Schwinger effect in the Salpeter equation.

\begin{figure}[t]
    \centering
    \includegraphics[width=0.35\linewidth]{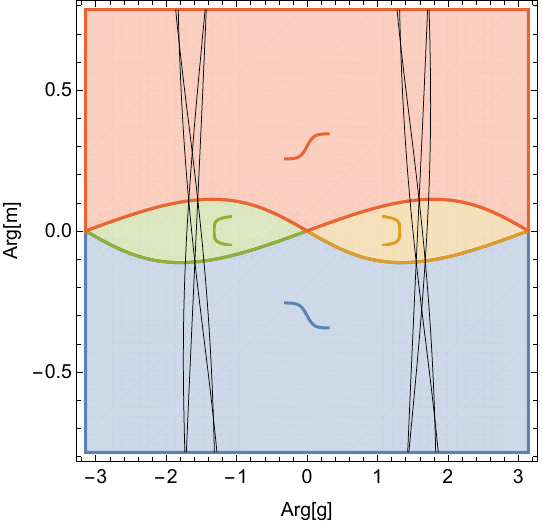}
    \caption{Stokes lines associated with the divergent saddle $B_-^{(0)}$ for $|m|=5$, $|g|=0.1$, and $x=350$. The boundaries of the colored regions denote Stokes lines for the degeneracy of $B_-^{(0)}$ and $B_+^{(0,-1)}$, which alter the topology of the $B_-^{(0)}$ thimble in the $z$-plane as indicated. Black lines indicate potential Stokes lines arising from the degeneracy of $B_-^{(0)}$ with $S_{\pm}^{(0,-1)}$. The illustrated range $|\Im(m)| < \pi/4$ captures the characteristic structure that remains invariant for $|\Im(m)| < \pi/2$, beyond which the roles of $B_-^{(0)}$ and $B_+^{(0)}$ are interchanged. In the large-$x$ limit, the black lines become increasingly vertical and the green and orange regions contract.}   \label{fig:complex}
\end{figure}

Notably, this mechanism remains robust even for complex $g$ and $m$ parameters, a regime where the relativistic $\Ai$ functions diverge as $e^{\pm i g x^2}$ and consequently cease to belong to $\mathcal{S}'(\mathbb{R})$.
Fig.~\ref{fig:complex} illustrates the Stokes lines associated with $B_-^{(0)}$ at a large positive $x$ in the $\arg(g)$-$\arg(m)$ plane.
In the vicinity of real values, we observe two Stokes lines for the degeneracy of $B_-^{(0)}$ and $B_+^{(0,-1)}$.
Crossing these lines induces a topological transition in the $B_-^{(0)}$ thimble, which is modified by the addition or subtraction of the $B_+^{(0,-1)}$ thimble.
However, since the $B_+^{(0,-1)}$ thimbles are horizontal, the global interconnectivity remains governed by $B_-^{(0)}$.
Meanwhile, the Stokes lines for the degeneracy of $B_-^{(0)}$ and $S_{\pm}^{(0,-1)}$ appear near $\arg(g) \simeq \pm\pi/2$ and become vertical in the limit $x \to \infty$. This behavior is elucidated by Eqs.~\eqref{eq:saddle_S} and \eqref{eq:saddle_B}, which indicate that the degeneracy of the imaginary parts is attained only for $\arg(g) \simeq \pm\pi/2$ at large $x$.
Crossing these Stokes lines causes the $B_-^{(0)}$ thimble to connect a more immediate pair of steepest descent regions. Regardless, the suppression of global interconnectivity remains robust for parameter values in the vicinity of the real axis. An analogous argument holds for the limit of large negative $x$.

\section{Summary}\label{sec:summary}
In this work, we presented a comprehensive analysis of the Salpeter, Dirac, and Klein-Gordon equations under a strong electric field using Picard-Lefschetz theory. Our primary objective was to provide a geometric explanation for the absence of Klein tunneling and the Schwinger effect in the Salpeter equation, a phenomenon that contrasts sharply with its presence in the Dirac and Klein-Gordon equations.

We demonstrated that the non-local square-root operator in the Salpeter equation necessitates a treatment within the framework of algebraic analysis, where solutions are constructed from pairings of integral kernels and fast-decay cycles on a Riemann surface.
Within the framework of the Riemann-Hilbert correspondence, this approach demonstrates that the general solutions to the Salpeter equation constitute a four-dimensional space. This solution space encompasses both positive- and negative-energy components, represented by relativistic generalizations of the Airy $\Ai$ and $\Bi$ functions and their respective negative-energy analogues.
These results are consistent with the non-relativistic limit, and suggest the applicability of the Riemann-Hilbert correspondence within the context of the system analyzed herein.

A key finding emerged from the comparison of Lefschetz thimble structures in the $z$-coordinate, where the exponents are fundamentally distinguished by the presence of branch points.
For the Dirac and Klein-Gordon equations, the thimbles associated with saddle points $S_-^{(n)}$ exhibit inter-period connectivity, bridging steepest descent directions across adjacent periods. This connectivity is a geometric prerequisite for Klein tunneling, allowing for the particle-antiparticle pair creation characteristic of the Schwinger effect. We successfully derived the Bogoliubov coefficients and pair production rates for these equations, consistent with established results.

In stark contrast, the Lefschetz thimbles for the Salpeter equation, particularly those originating from $S_-^{(n)}$, do not connect across different periods. Instead, divergent branch points ($B_\pm^{(n)}$) have inter-period connectivity and are inherent to the non-local operator.
Once we avoid these divergent branch points, inter-period connectivity is lost, which geometrically restricts the admissible solution cycles to regions that prevent the necessary transitions for pair creation. Consequently, only the relativistic $\Ai$ functions, which remain bounded, constitute physical solutions, effectively blocking the Schwinger effect.
This robustness, extending to complex parameters, underscores the fundamental nature of this geometric interpretation for the absence of the Schwinger effect in the Salpeter equation.

Our analysis provides a unified geometric interpretation of Klein tunneling and the Schwinger effect across different relativistic wave equations, demonstrating that Lefschetz thimble analysis offers a more comprehensive perspective on the analysis of physical systems.
Furthermore, it highlights the profound implications of the non-local nature of the Salpeter operator and its associated Riemann-Hilbert correspondence for irregular holonomic $\mathcal{E}$-modules, offering insights into solution space of such systems.

\begin{acknowledgments}
    The author is supported by the Slovenian Research Agency under the research grant J1-4389. The author thanks Gianluca Lagnese for the fruitful discussion that motivated the calculation of the relativistic $\Bi$ function.
\end{acknowledgments}

\bibliographystyle{apsrev4-1}
\bibliography{airy}

@article{PhysRev.82.664,
  title = {On Gauge Invariance and Vacuum Polarization},
  author = {Schwinger, Julian},
  journal = {Phys. Rev.},
  volume = {82},
  issue = {5},
  pages = {664--679},
  numpages = {0},
  year = {1951},
  month = {Jun},
  publisher = {American Physical Society},
  doi = {10.1103/PhysRev.82.664},
  url = {https://link.aps.org/doi/10.1103/PhysRev.82.664}
}

@article{Sauter:1931zz,
    author = "Sauter, Fritz",
    title = "{Uber das Verhalten eines Elektrons im homogenen elektrischen Feld nach der relativistischen Theorie Diracs}",
    doi = "10.1007/BF01339461",
    journal = "Z. Phys.",
    volume = "69",
    pages = "742--764",
    year = "1931"
}

@article{Weisskopf:406571,
      author        = "Weisskopf, Victor Frederick",
      title         = "{Über die Elektrodynamik des Vakuums auf Grund des
                       Quanten-Theorie des Elektrons}",
      journal       = "Dan. Mat. Fys. Medd.",
      volume        = "14",
      number        = "6",
      pages         = "1-39",
      year          = "1936",
      url           = "https://cds.cern.ch/record/406571",
}

@article{Heisenberg:1936nmg,
    author = "Heisenberg, W. and Euler, H.",
    title = "{Consequences of Dirac's theory of positrons}",
    eprint = "physics/0605038",
    archivePrefix = "arXiv",
    doi = "10.1007/BF01343663",
    journal = "Z. Phys.",
    volume = "98",
    number = "11-12",
    pages = "714--732",
    year = "1936"
}

@article{10.1119/1.1934851,
    author = {Winter, Rolf G.},
    title = {Klein Paradox for the Klein-Gordon Equation},
    journal = {American Journal of Physics},
    volume = {27},
    number = {5},
    pages = {355-358},
    year = {1959},
    month = {05},
    issn = {0002-9505},
    doi = {10.1119/1.1934851},
    url = {https://doi.org/10.1119/1.1934851}
}

@article{1929ZPhy...53..157K,
       author = {{Klein}, O.},
        title = "{Die Reflexion von Elektronen an einem Potentialsprung nach der relativistischen Dynamik von Dirac}",
      journal = {Zeitschrift fur Physik},
         year = 1929,
        month = mar,
       volume = {53},
       number = {3-4},
        pages = {157-165},
          doi = {10.1007/BF01339716},
       adsurl = {https://ui.adsabs.harvard.edu/abs/1929ZPhy...53..157K}
}

@article{10.1063/1.530015,
    author = {Lämmerzahl, Claus},
    title = {The pseudodifferential operator square root of the Klein-Gordon equation},
    journal = {Journal of Mathematical Physics},
    volume = {34},
    number = {9},
    pages = {3918-3932},
    year = {1993},
    month = {09},
    issn = {0022-2488},
    doi = {10.1063/1.530015},
    url = {https://doi.org/10.1063/1.530015}
}

@article{Daem:2024xrc,
    author = "Daem, F. and Matzkin, A.",
    title = {{Tunneling dynamics of the relativistic Schr{\"o}dinger/Salpeter equation}},
    eprint = "2406.16644",
    archivePrefix = "arXiv",
    primaryClass = "quant-ph",
    doi = "10.1088/1402-4896/ad9550",
    journal = "Phys. Scripta",
    volume = "100",
    number = "1",
    pages = "015216",
    year = "2025"
}

@article{Zumer:2025jqh,
    author = "Zumer, Beno{\^\i}t and Daem, Florent and Matzkin, Alexandre",
    title = {{Revivals and quantum carpets for the relativistic Schr{\"o}dinger equation}},
    eprint = "2511.05200",
    archivePrefix = "arXiv",
    primaryClass = "quant-ph",
    doi = "10.1016/j.physleta.2026.131549",
    journal = "Phys. Lett. A",
    volume = "582",
    pages = "131549",
    year = "2026"
}

@article{Kowalski:2011zi,
    author = "Kowalski, K. and Rembielinski, J.",
    title = "{The Salpeter equation and probability current in the relativistic Hamiltonian quantum mechanics}",
    eprint = "1110.5146",
    archivePrefix = "arXiv",
    primaryClass = "math-ph",
    doi = "10.1103/PhysRevA.84.012108",
    journal = "Phys. Rev. A",
    volume = "84",
    pages = "012108",
    year = "2011"
}

@article{Akhmedov:2020dgc,
    author = "Akhmedov, E. T. and Anokhin, A. V. and Sadekov, D. I.",
    title = "{Currents of created pairs in strong electric fields}",
    eprint = "2012.00399",
    archivePrefix = "arXiv",
    primaryClass = "hep-th",
    doi = "10.1142/S0217751X21501347",
    journal = "Int. J. Mod. Phys. A",
    volume = "36",
    number = "19",
    pages = "2150134",
    year = "2021"
}

@article{Allen:2003wz,
    author = "Allen, Theodore J. and Olsson, M. G.",
    title = "{Reduction of the QCD string to a time component vector potential}",
    eprint = "hep-ph/0306128",
    archivePrefix = "arXiv",
    reportNumber = "MADPH-03-1324",
    doi = "10.1103/PhysRevD.68.054022",
    journal = "Phys. Rev. D",
    volume = "68",
    pages = "054022",
    year = "2003"
}

@article{Buisseret:2006sz,
    author = "Buisseret, Fabien and Mathieu, Vincent",
    title = "{Hybrid mesons and auxiliary fields}",
    eprint = "hep-ph/0607083",
    archivePrefix = "arXiv",
    doi = "10.1140/epja/i2006-10090-0",
    journal = "Eur. Phys. J. A",
    volume = "29",
    pages = "343--351",
    year = "2006"
}

@article{PhysRevB.60.14525,
  title = {Decay of the metastable phase in $d=1$ and $d=2$ Ising models},
  author = {Rutkevich, S. B.},
  journal = {Phys. Rev. B},
  volume = {60},
  issue = {21},
  pages = {14525--14528},
  numpages = {0},
  year = {1999},
  month = {Dec},
  publisher = {American Physical Society},
  doi = {10.1103/PhysRevB.60.14525},
  url = {https://link.aps.org/doi/10.1103/PhysRevB.60.14525}
}

@ARTICLE{2019arXiv190801276D,
       author = {{D'Agnolo}, Andrea and {Kashiwara}, Masaki},
        title = "{Enhanced specialization and microlocalization}",
      journal = {arXiv e-prints},
         year = 2019,
        month = aug,
          eid = {arXiv:1908.01276},
        pages = {arXiv:1908.01276},
          doi = {10.48550/arXiv.1908.01276},
archivePrefix = {arXiv},
       eprint = {1908.01276},
 primaryClass = {math.AG},
       adsurl = {https://ui.adsabs.harvard.edu/abs/2019arXiv190801276D}
}

@article{Waschkies2005MicrolocalRC,
  title={Microlocal Riemann-Hilbert Correspondence},
  author={Ingo Waschkies},
  journal={Publications of The Research Institute for Mathematical Sciences},
  year={2005},
  volume={41},
  pages={37-72},
  url={https://api.semanticscholar.org/CorpusID:54683090}
}

@article{CM_1984__51_1_51_0,
     author = {Mebkhout, Z.},
     title = {Une \'equivalence de cat\'egories},
     journal = {Compositio Mathematica},
     pages = {51--62},
     year = {1984},
     publisher = {Martinus Nijhoff Publishers},
     volume = {51},
     number = {1},
     mrnumber = {734784},
     zbl = {0566.32021},
     language = {fr},
     url = {https://www.numdam.org/item/CM_1984__51_1_51_0/}
}

@article{Kashiwara1984,
  author    = {Kashiwara, Masaki},
  title     = {The Riemann-Hilbert problem for holonomic systems},
  journal   = {Publications of the Research Institute for Mathematical Sciences},
  volume    = {20},
  number    = {2},
  pages     = {319--365},
  year      = {1984},
  doi       = {10.2977/prims/1195181646}
}

@article{PhysRev.87.328,
  title = {Mass Corrections to the Fine Structure of Hydrogen-Like Atoms},
  author = {Salpeter, E. E.},
  journal = {Phys. Rev.},
  volume = {87},
  issue = {2},
  pages = {328--343},
  numpages = {0},
  year = {1952},
  month = {Jul},
  publisher = {American Physical Society},
  doi = {10.1103/PhysRev.87.328},
  url = {https://link.aps.org/doi/10.1103/PhysRev.87.328}
}

@article{Gavrilov:2015zem,
    author = "Gavrilov, S. P. and Gitman, D. M.",
    title = "{Scattering and pair creation by a constant electric field between two capacitor plates}",
    eprint = "1511.02915",
    archivePrefix = "arXiv",
    primaryClass = "hep-th",
    doi = "10.1103/PhysRevD.93.045033",
    journal = "Phys. Rev. D",
    volume = "93",
    number = "4",
    pages = "045033",
    year = "2016"
}

@article{Gavrilov:2007hq,
    author = "Gavrilov, S. P. and Gitman, Dmitry M.",
    title = "{One-loop energy-momentum tensor in QED with electric-like background}",
    eprint = "0709.1828",
    archivePrefix = "arXiv",
    primaryClass = "hep-th",
    reportNumber = "PUBLICACAO-IF-USP-1641-2007",
    doi = "10.1103/PhysRevD.78.045017",
    journal = "Phys. Rev. D",
    volume = "78",
    pages = "045017",
    year = "2008"
}

@inproceedings{Ecalle1981LesFR,
  title={Les fonctions r{\'e}surgentes appliqu{\'e}es {\`a} l'it{\'e}ration},
  author={Jean Ecalle},
  year={1981},
  url={https://api.semanticscholar.org/CorpusID:118298486}
}

@inproceedings{Kawai2005AlgebraicAO,
  title={Algebraic Analysis of Singular Perturbation Theory},
  author={Takahiro Kawai and Yoshitsugu Takei},
  year={2005},
  url={https://api.semanticscholar.org/CorpusID:116690639}
}

@article{Voros1983TheRO,
  title={The return of the quartic oscillator. The complex WKB method},
  author={Andr{\'e} Voros},
  journal={Annales De L Institut Henri Poincare-physique Theorique},
  year={1983},
  volume={39},
  pages={211-338},
  url={https://api.semanticscholar.org/CorpusID:117405396}
}

@article{Witten:2010cx,
    author = "Witten, Edward",
    editor = "Andersen, Joergen E. and Boden, Hans U. and Hahn, Atle and Himpel, Benjamin",
    title = "{Analytic Continuation Of Chern-Simons Theory}",
    eprint = "1001.2933",
    archivePrefix = "arXiv",
    primaryClass = "hep-th",
    journal = "AMS/IP Stud. Adv. Math.",
    volume = "50",
    pages = "347--446",
    year = "2011"
}

@article{Hansen_1981,
doi = {10.1088/0031-8949/23/6/002},
url = {https://doi.org/10.1088/0031-8949/23/6/002},
year = {1981},
month = {jun},
publisher = {},
volume = {23},
number = {6},
pages = {1036},
author = {Alex Hansen and Finn Ravndal},
title = {Klein's Paradox and Its Resolution},
journal = {Physica Scripta}
}

\appendix
\section{Relativistic Airy functions}\label{apx:relativistic-airy}
This appendix provides a detailed analysis of the relativistic Airy functions.
\subsection{Cycles for the relativistic Airy functions}
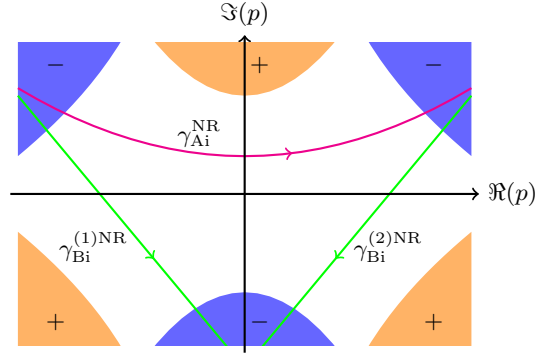
\begin{figure}
    \centering
    \begin{tikzpicture}
        \begin{scope}
        \clip (-3,-2) -- (3,-2) -- (3,2) -- (-3,2) -- cycle;
        \fill[orange!60] plot[domain=-3:-0.1] (\x, {3/\x-0.5*\x-1}) -- (-3,-3) -- cycle;
        \fill[orange!60] plot[domain=0.1:3] (\x, {-3/\x+0.5*\x-1}) -- (3,-3) -- cycle;
        \fill[orange!60] plot[domain=-2:2] (\x, {0.5*\x*\x+1.3}) -- cycle;
        \fill[blue!60] plot[domain=-3:-0.1] (\x, {-3/\x+0.5*\x+1}) -- (-3,3) -- cycle;
        \fill[blue!60] plot[domain=0.1:3] (\x, {3/\x-0.5*\x+1}) -- (3,3) -- cycle;
        \fill[blue!60] plot[domain=-2:2] (\x, {-0.5*\x*\x-1.3}) -- cycle;
        \draw[thick, magenta, style={decoration={markings, mark=at position 0.6 with {\arrow{>}}},postaction={decorate}}] plot[domain=-3:3] (\x, {0.1*\x*\x+0.5});
        \draw[thick, green, style={decoration={markings, mark=at position 0.6 with {\arrow{>}}},postaction={decorate}}] plot[domain=-3:0] (\x, {-1.2*\x-2.3});
        \draw[thick, green, style={decoration={markings, mark=at position 0.6 with {\arrow{>}}},postaction={decorate}}] plot[domain=3:0] (\x, {1.2*\x-2.3});
        \end{scope}
        \draw[->, thick] (-3.1,0) -- (3.1,0) node[right] {$\Re(p)$};
        \draw[->, thick] (0,-2.1) -- (0,2.1) node[above] {$\Im(p)$};
        \node at (0.2,1.7) {$+$};
        \node at (-2.5,1.7) {$-$};
        \node at (2.5,1.7) {$-$};
        \node at (2.5,-1.7) {$+$};
        \node at (-2.5,-1.7) {$+$};
        \node at (0.2,-1.7) {$-$};
        \node at (-0.6,0.8) {$\gamma_{\Ai}^{\rm NR}$};
        \node at (-2,-0.7) {$\gamma_{\Bi}^{(1)\rm NR}$};
        \node at (1.9,-0.7) {$\gamma_{\Bi}^{(2)\rm NR}$};
    \end{tikzpicture}
    \caption{Schematic illustration of the steepest ascent (orange) and descent (blue) directions for the Airy equation.}
    \label{fig:airy}
\end{figure}

This subsection defines the relativistic Airy functions by comparing their integral representations with those of the classical Airy functions.
In the non-relativistic limit ($m \to \infty$), the Salpeter equation~\eqref{eq:salpeter} reduces to
\begin{equation}
    0=\ab[\sqrt{-\partial^2+m^2}+gx]f(x)\simeq\frac{1}{2m}\ab[-\partial^2+2mg\ab(x+\frac{m}{g})]f(x).
\end{equation}
Introducing $F(x)=f(x-m/g)$ and setting $g=1/(2m)$, the $m\to\infty$ limit yields the differential equation
\begin{equation}
    (-\partial^2+x)F(x)= 0,
\end{equation}
which is the Airy equation.
This equation admits two linearly independent solutions, represented by the following integral forms:
\begin{align}
    \Ai(x)&=\int_{\gamma_{\Ai}^{\rm NR}}\frac{\odif{p}}{2\pi}e^{i\frac{p^3}{3}+ipx},\label{eq:airy_integral_ai}\\
    \Bi(x)&=i\int_{\gamma_{\Bi}^{(1)\rm NR}}\frac{\odif{p}}{2\pi}e^{i\frac{p^3}{3}+ipx}+i\int_{\gamma_{\Bi}^{(2)\rm NR}}\frac{\odif{p}}{2\pi}e^{i\frac{p^3}{3}+ipx},\label{eq:airy_integral_bi}
\end{align}
where the cycles are indicated in Fig.~\ref{fig:airy}.

Let us return to the Salpeter equation with $g=1/(2m)$, namely
\begin{equation}
    \ab[2m\ab(\sqrt{-\partial^2+m^2}-m)+x]F(x)=0.
\end{equation}
In the limit $m \gg |p|$, the exponent admits the expansion
\begin{align}
    \mathcal I_{S}(x-m/g,p)|_{g=1/(2m)}&=i\frac{p^3}{3}+ipx+\order\ab(\frac{p^5}{m^2}),
\end{align}
which matches the exponents in Eqs.~\eqref{eq:airy_integral_ai} and \eqref{eq:airy_integral_bi}. Consequently, the integration cycles for the relativistic Airy functions must be analogous for $|p| \ll m$. The size of the contribution from the integration over the region $|p| \gtrsim m$ depends on the saddle-point locations. If the saddles are situated well within $|p| < m$, the integrand is strongly suppressed, allowing large-$|p|$ contributions to be neglected. Conversely, if the saddles are not contained, the functions are affected by these contributions, as expected since relativistic effects emerge when the classical momentum exceeds $m$.
Thus, the non-relativistic limit is valid specifically in the region where $p \sim |x/(2m)| \ll m$.

The cycle for the relativistic $\Ai$ function is straightforwardly identified because the associated steepest descent directions coincide with the non-relativistic case. For the relativistic $\Bi$ function, the cycle must traverse the branch cut, and we adopt the simplest cycles for our definition.
These cycles are illustrated in Fig.~\ref{fig:sigma}.
Accordingly, we define the relativistic Airy functions as follows:
\begin{align}
    \Ai_{m}(x)&=\left.\int_{\gamma_{\Ai}}\frac{\odif{p}}{2\pi}e^{\mathcal I_{S}(x-m/g,p)}\right|_{g=1/(2m)},\\
    \Bi_{m}(x)&=\left.i\int_{\gamma_{\Bi}^{(1)}}\frac{\odif{p}}{2\pi}e^{\mathcal I_{S}(x-m/g,p)}+i\int_{\gamma_{\Bi}^{(2)}}\frac{\odif{p}}{2\pi}e^{\mathcal I_{S}(x-m/g,p)}\right|_{g=1/(2m)}.
\end{align}
Specifically, the relativistic $\Bi$ function necessitates the inclusion of both the positive and negative Riemann sheets; this requirement for multiple sheets explains why the function cannot be derived using the standard Fourier transform method.

\begin{figure}[t]
    \centering
    \includegraphics[width=0.4\linewidth]{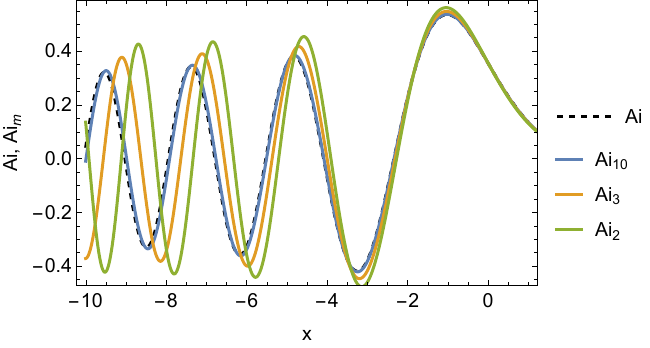}\quad
    \includegraphics[width=0.4\linewidth]{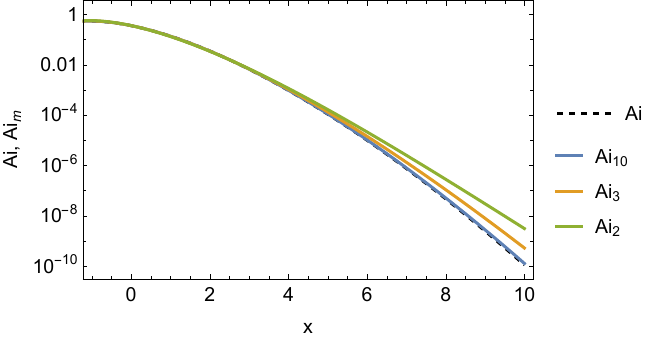}
    \includegraphics[width=0.4\linewidth]{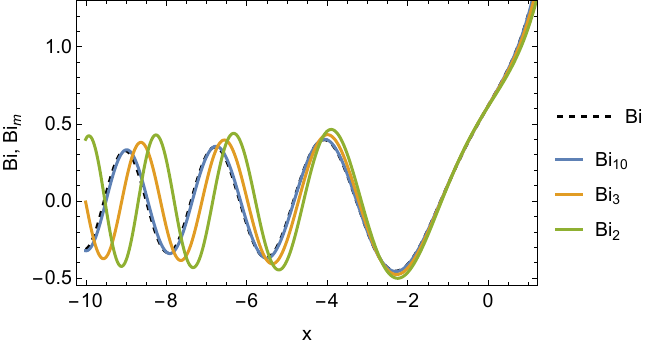}\quad
    \includegraphics[width=0.4\linewidth]{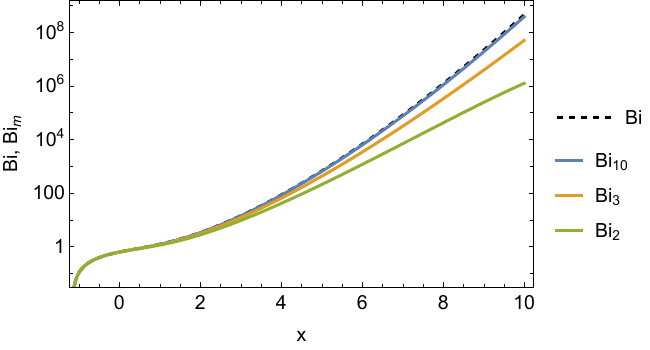}
    \caption{The relativistic Airy functions (solid) and the classical Airy functions (dashed). The values $m=2, 3, 10$ are used for illustrative purposes.}
    \label{fig:shape_functions}
\end{figure}

Fig.~\ref{fig:shape_functions} compares the relativistic Airy functions with the classical Airy functions.
Note that these functions are real by construction.
While they approach the classical Airy functions as $m \to \infty$, the convergence is non-uniform because the classical momentum exceeds $m$ for sufficiently large $|x|$. As $m$ decreases, the oscillation frequency and the amplitude for $x < 0$ increase, and the exponential growth and decay for $x > 0$ become less pronounced.
As we will see, the relativistic $\Bi$ function becomes negative for sufficiently large $x$, implying the existence of an additional zero in the region $x > 0$.

\subsection{Stokes lines and asymptotic behaviors}
We now examine the Stokes lines for the relativistic Airy functions in comparison with those of the classical Airy functions. Let us first summarize the saddle points.
Following the reparametrization in the previous subsection, the saddle points $S_\pm^{(0)}$ for the Salpeter equation are given by
\begin{equation}
    p=\pm\sqrt{-x\ab(1-\frac{x}{4m^2})}.
\end{equation}
Notice that the sign conventions coincide only for $x < 2m^2$. For the classical Airy equation, the saddle points are given by
\begin{equation}
    p_{S_\pm}=\pm\sqrt{-x}.
\end{equation}
We see that the saddle points of the Salpeter equation coincide with those of the Airy equation in the limit $|x| \ll 4m^2$. Note that the square root introduces a branch cut along the positive real axis, across which the two saddle points $p_{S_\pm}$ are interchanged.

\begin{figure}[t]
    \centering
    \includegraphics[width=0.305\linewidth]{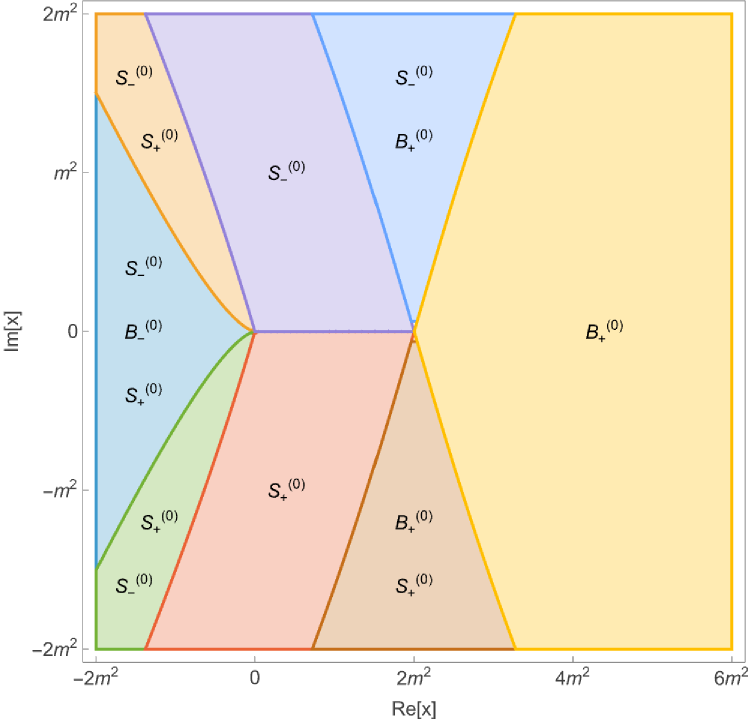}\quad
    \includegraphics[width=0.305\linewidth]{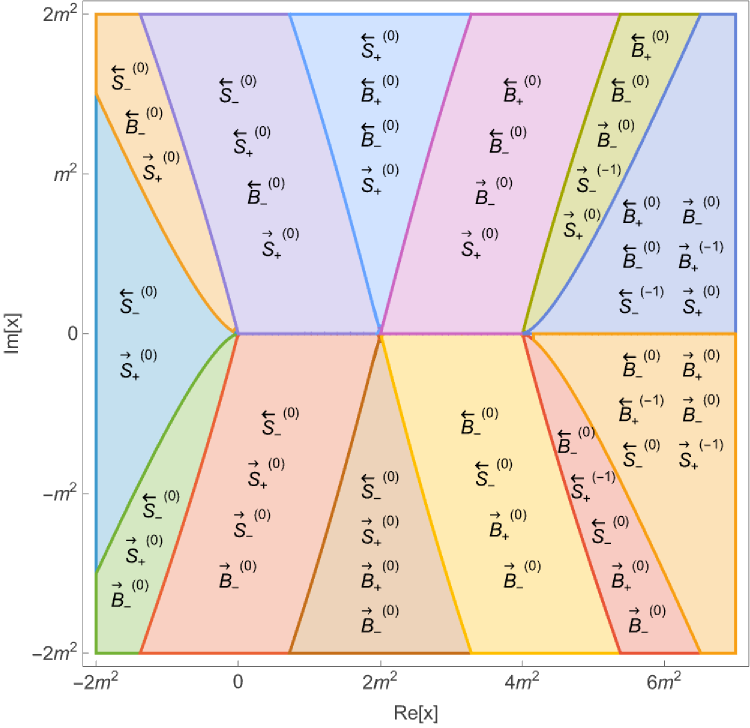}\quad
    \includegraphics[width=0.29\linewidth]{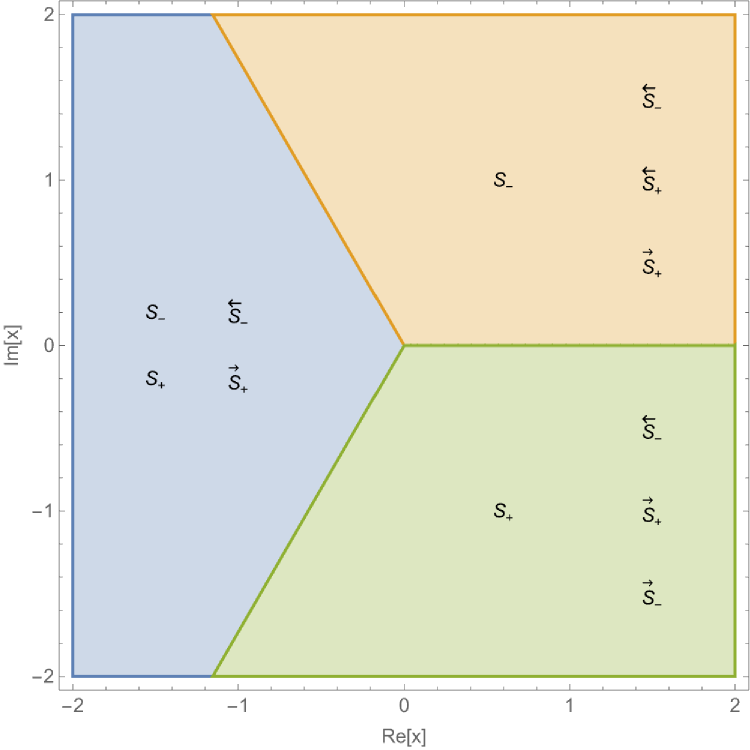}
    \caption{The stokes lines for the relativistic $\Ai$ function (left), relativistic $\Bi$ function (middle) and the classical Airy functions (right). The left arrows indicate the contributing saddles for $\gamma_{\Bi}^{(1)}$ and the right arrows indicate those for $\gamma_{\Bi}^{(2)}$.}
    \label{fig:stokes_lines}
\end{figure}

Fig.~\ref{fig:stokes_lines} illustrates the Stokes lines for the relativistic and classical Airy functions. Neglecting the subdominant branch points $B_{-}^{(n)}$, the topology of the Stokes phenomena aligns with that of the classical Airy function in regions 1, 2, 3, 1', 2', and 3' in Fig.~\ref{fig:stokes}. For $x < 0$, the differences are limited to the relocations of the saddle points and corrections to the exponent. However, for $x > 2m^2$, a fundamental difference arises by crossing the Stokes lines, leading to distinct asymptotic behaviors as $x \to \infty$.

To evaluate the asymptotic behavior, a refined treatment of the saddle-point approximation around $B_{\pm}^{(n)}$ is required, as the prefactor is an odd function. This property results in a leading-order cancellation of the contributions from symmetric regions around the saddle point. This cancellation is only partial, however, with the residual contribution arising from higher-order terms in the expansion of the exponent.
Specifically, in the vicinity of the branch points, the integrand admits the following expansion:
\begin{equation}
    m\cosh(z)e^{\mathcal I_S(x-m/g,z)|_{g=1/(2m)}} \simeq \pm im\ab(z-z_{B_\pm^{(0)}})\ab(1-i\frac{2m^3}{3}\ab(z-z_{B_\pm^{(0)}})^3)e^{\mp(mx-\frac{m^3}{2}(4-\pi))\mp\frac{mx}{2}\ab(z-z_{B_\pm^{(0)}})^2},
\end{equation}
where the cubic term in the exponent has been expanded.

In the $x \to \infty$ limit, the asymptotic expansions for the $\Ai$ functions are
\begin{align}
    \Ai(x)&\simeq\frac{1}{2\sqrt{\pi}}\frac{1}{x^{\frac14}}e^{-\frac{2}{3}x^{\frac32}},\\
    \Ai_{m}(x)&\simeq \sqrt{\frac{2m}{\pi}}\frac{m}{x^{\frac{5}{2}}}e^{-mx+\frac{m^3}{2}(4-\pi)}.
\end{align}
While both functions exhibit decay, the leading exponential contribution to the relativistic $\Ai$ function originates from a branch point rather than a saddle point.

For the $\Bi$ functions, the positive real axis lies on a Stokes line, and the thimble configuration depends on the sign of $\Im(x)$. For $\Im(x) > 0$, the expansions are given by
\begin{align}
    \Bi(x)&\simeq\frac{1}{\sqrt{\pi}}\frac{1}{x^{\frac14}}e^{\frac{2}{3}x^{\frac32}}+i\frac{1}{2\sqrt{\pi}}\frac{1}{x^{\frac14}}e^{-\frac{2}{3}x^{\frac32}},\\
    \Bi_{m}(x)&\simeq -\sqrt{\frac{2m}{\pi}}\frac{2m}{x^{\frac{5}{2}}}e^{mx-\frac{m^3}{2}(4-\pi)}+\frac{e^{m^3\pi}}{\sqrt{\pi m}}\Im\ab(e^{i\ab(-\frac{x^2}{4m}+mx-\frac{m^3}{2}+\frac{\pi}{4})}\ab(\frac{m^2}{x})^{-im^3})
    \nonumber\\
    &\hspace{3ex}+i\ab(1+e^{2m^3\pi})\sqrt{\frac{2m}{\pi}}\frac{m}{x^{\frac{5}{2}}}e^{-mx+\frac{m^3}{2}(4-\pi)}.
\end{align}
Note that the leading term of the relativistic $\Bi$ function is negative, indicating that the function undergoes a sign change at some point in the region $x > 0$.

Conversely, for $\Im(x) < 0$, they are given by
\begin{align}
    \Bi(x)&\simeq\frac{1}{\sqrt{\pi}}\frac{1}{x^{\frac14}}e^{\frac{2}{3}x^{\frac32}}-i\frac{1}{2\sqrt{\pi}}\frac{1}{x^{\frac14}}e^{-\frac{2}{3}x^{\frac32}},\\
    \Bi_{m}(x)&\simeq -\sqrt{\frac{2m}{\pi}}\frac{2m}{x^{\frac{5}{2}}}e^{mx-\frac{m^3}{2}(4-\pi)}+\frac{e^{m^3\pi}}{\sqrt{\pi m}}\Im\ab(e^{i\ab(-\frac{x^2}{4m}+mx-\frac{m^3}{2}+\frac{\pi}{4})}\ab(\frac{m^2}{x})^{-im^3})
    \nonumber\\
    &\hspace{3ex}-i\ab(1+e^{2m^3\pi})\sqrt{\frac{2m}{\pi}}\frac{m}{x^{\frac{5}{2}}}e^{-mx+\frac{m^3}{2}(4-\pi)}.
\end{align}
Small imaginary components appear in the expansions of both $\Bi$ and $\Bi_m$, with signs determined by $\Im(x)$. This is a well-known phenomenon in saddle-point approximations, where a spurious imaginary part arises despite the functions being strictly real. 
Notably, the relativistic $\Bi$ function grows at a slower rate determined by the branch point and includes additional oscillatory contributions analogous to Klein tunneling.

In the $x \to -\infty$ limit, the asymptotic behaviors are
\begin{align}
    \Ai(x)&\simeq\frac{1}{\sqrt{\pi}}\Re\ab(\frac{1}{(-x)^{\frac14}}e^{i\ab(-\frac{2}{3}(-x)^{\frac32}+\frac{\pi}{4})}),\\
    \Ai_{m}(x)&\simeq \frac{1}{\sqrt{\pi m}}\Re\ab(e^{i\ab(-\frac{x^2}{4m}+mx-\frac{m^3}{2}+\frac{\pi}{4})}\ab(\frac{m^2}{-x})^{-im^3})-\sqrt{\frac{2m}{\pi}}\frac{m}{(-x)^{\frac{5}{2}}}e^{mx-\frac{m^3}{2}(4-\pi)},\\
    \Bi(x)&\simeq\frac{1}{\sqrt{\pi}}\Im\ab(\frac{1}{(-x)^{\frac14}}e^{i\ab(-\frac{2}{3}(-x)^{\frac32}+\frac{\pi}{4})}),\\
    \Bi_{m}(x)&\simeq \frac{1}{\sqrt{\pi m}}\Im\ab(e^{i\ab(-\frac{x^2}{4m}+mx-\frac{m^3}{2}+\frac{\pi}{4})}\ab(\frac{m^2}{-x})^{-im^3}).
\end{align}
For all functions, the leading contributions originate from saddle points, not from branch points.
We observe that the relativistic Airy functions exhibit higher-frequency oscillations. The relativistic $\Ai$ function also possesses a subdominant, exponentially damped contribution, which stems from a branch point.

\section{Breakdown of saddle point approximation}\label{apx:breakdown}
In the Dirac and Klein-Gordon equations, the standard saddle-point approximation is rendered inadequate for evaluating the contributions from the $S_-^{(n)}$ thimbles in the regime where $m^2/g$ is small. This breakdown is attributed to the Lefschetz thimble's proximity to an adjacent saddle point, resulting in a significant deviation from the local Gaussian distribution.
This appendix presents a numerical investigation of this phenomenon.

\begin{figure}[t]
    \centering
    \includegraphics[width=0.35\linewidth]{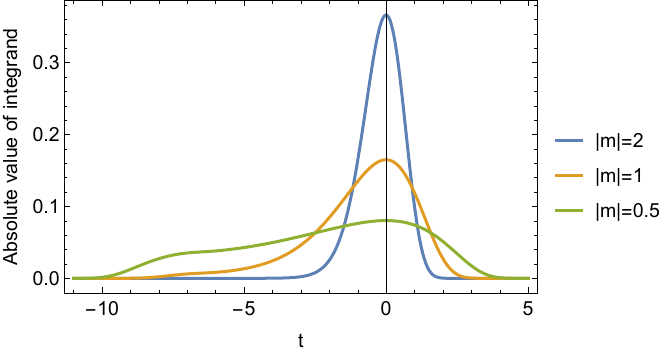}\quad
    \includegraphics[width=0.2\linewidth]{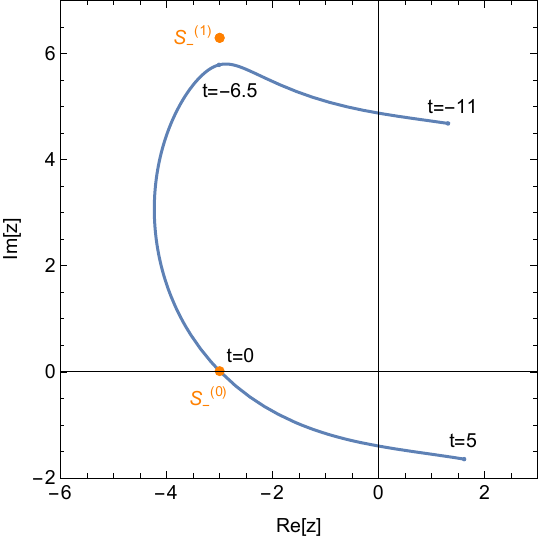}
    \caption{The integrand for the lower component of the Dirac spinor along the Lefschetz thimble associated with $S^{(0)}_-$ for various values of $|m|$. Parameters are set to $g=1$, $\arg(m)=0.01$, and $x=-10|m|/g$. The right panel shows the Lefschetz thimbles, which are common for all values of $|m|$.}
    \label{fig:non-gaussian}
\end{figure}

Fig.~\ref{fig:non-gaussian} illustrates the integrand for the lower component of the Dirac spinor, $\frac{m}{2\pi}\odv{z}{t}e^{\mathcal I_{DKG}}$, evaluated along the $S_-^{(0)}$ thimble $z(t)$ for various values of $|m|$ with fixed $g=1$ and $x=-10|m|/g$. We set $\arg(m)=0.01$ to resolve degenerated saddles. Here, the parameter $t$ denotes the arc length along the Lefschetz thimble, measured from the saddle point.
While the integrand exhibits a Gaussian profile for large $|m|$, it develops an extended, non-trivial structure as $|m|$ decreases, reflecting the influence of adjacent saddle points. The right panel displays the Lefschetz thimble, which remains invariant with respect to $|m|$. As evidenced by the thimble's trajectory, the path from $S_-^{(0)}$ passes in close proximity to $S_-^{(1)}$ near $t \simeq -6.5$, resulting in a deviation from the local Gaussian approximation.

\begin{figure}[t]
    \centering
    \includegraphics[width=0.32\linewidth]{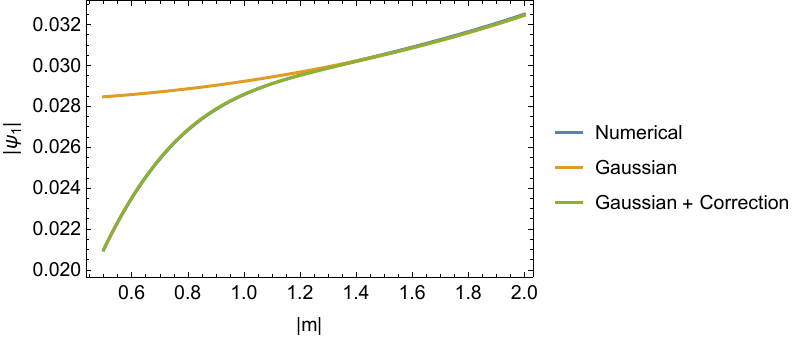}\quad
    \includegraphics[width=0.32\linewidth]{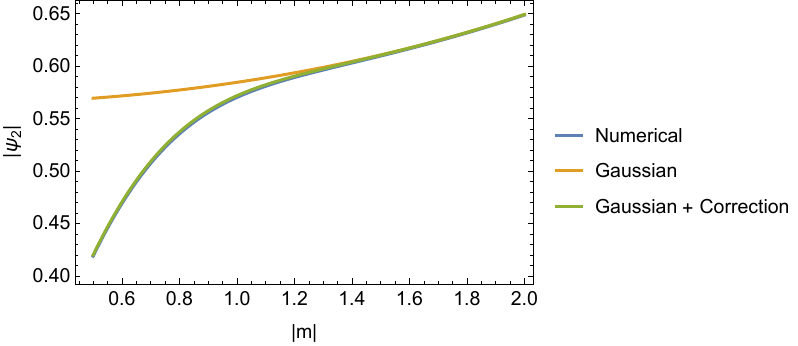}\quad
    \includegraphics[width=0.32\linewidth]{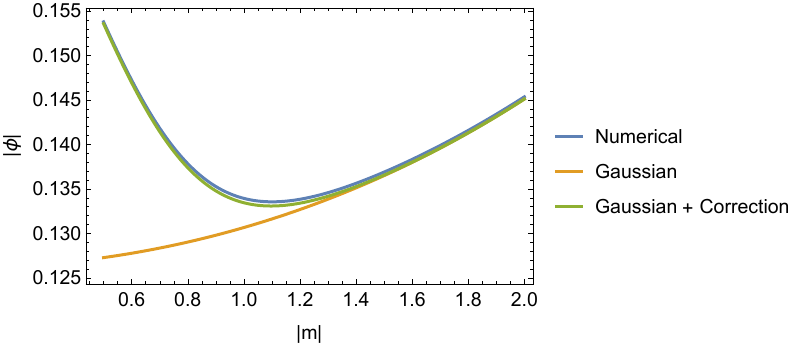}
    \caption{Comparison of integrated values along the Lefschetz thimble associated with $S^{(0)}_-$ for various values of $|m|$. The parameters and labels follow those established in Fig.~\ref{fig:non-gaussian}. Here, $\psi_1$ and $\psi_2$ denote the components of the Dirac spinor, while $\phi$ represents the Klein-Gordon scalar field.}
    \label{fig:numerical-test}
\end{figure}

As detailed in Sec.~\ref{sec:schwinger}, the requirement of current conservation provides a systematic way to determine the correction factors $|M_{\psi/\phi}|$ that account for this non-Gaussian behavior.
Fig.~\ref{fig:numerical-test} presents a comparison between direct numerical integration over the thimble and the leading-order saddle-point approximation, both with and without the correction factors $M_{\psi/\phi}$. The results demonstrate that the inclusion of $M_{\psi/\phi}$ accurately recovers the full integral values, validating the consistency of our current conservation approach in the regime where the standard saddle-point expansion fails.
Furthermore, numerical results indicate that this correction factor is equally applicable to the subdominant component of the Dirac spinor, $\psi_1$. Furthermore, our numerical results substantiate the sign difference between $M_\psi$ and $M_\phi$.

\section{Bogoliubov coefficients}\label{apx:Bogoliubov}
This appendix provides the derivation of the Bogoliubov coefficients necessary for the evaluation of the Schwinger effect.

We first define the positive-energy incoming and outgoing solutions. Note that a solution with energy eigenvalue $E$ is obtained by the shift $x \to x - E/g$.
In the region $x < 0$, the solutions correspond to particle states, and the incoming and outgoing solutions coincide with those defined in Sec.~\ref{sec:schwinger}.
In the region $x > 0$, however, the solutions characterize antiparticles with energy $-E$\footnote{One can readily verify that $\psi^c = C\bar{\psi}^T$ and $\phi^c = \phi^*$ satisfy the original differential equations under the substitutions $g \to -g$ and $E \to -E$.}.
Accordingly, the temporal evolution is inverted, leading to a reversal of the group velocity.
Subject to the normalization condition $|j_\psi(\pm\infty)| = 1$, the positive-energy solutions to the Dirac equation exhibit the following asymptotic behaviors:
\begin{align}
    \psi_{L,E}^{\rm in}(x)&\underset{x\to-\infty}{\simeq} e^{-i\frac{g}{2}x^2+iEx}\ab(-\frac{m}{2(gx-E)})^{-i\frac{m^2}{2g}}\begin{pmatrix}
        1\\
        -\frac{m}{2(gx-E)}
    \end{pmatrix},\label{eq:dirac_normalized_1}\\
    \psi_{L,E}^{\rm out}(x)&\underset{x\to-\infty}{\simeq} e^{i\frac{g}{2}x^2-iEx}\ab(-\frac{m}{2(gx-E)})^{i\frac{m^2}{2g}}\begin{pmatrix}
        -\frac{m}{2(gx-E)}\\
        1
    \end{pmatrix},\\
    \psi_{R,E}^{\rm in}(x)&\underset{x\to+\infty}{\simeq} e^{i\frac{g}{2}x^2+iEx}\ab(\frac{m}{2(gx+E)})^{i\frac{m^2}{2g}}\begin{pmatrix}
        -\frac{m}{2(gx+E)}\\
        1
    \end{pmatrix},\\
    \psi_{R,E}^{\rm out}(x)&\underset{x\to+\infty}{\simeq} e^{-i\frac{g}{2}x^2-iEx}\ab(\frac{m}{2(gx+E)})^{-i\frac{m^2}{2g}}\begin{pmatrix}
        1\\
        -\frac{m}{2(gx+E)}
    \end{pmatrix}.\label{eq:dirac_normalized_4}
\end{align}
A field operator $\hat\Psi(x)$ can then be expanded by these functions as
\begin{align}
    \hat \Psi(x)&=\int_0^\infty\odif{E}\ab[\psi_{L,E}^{\rm in}(x)\hat a_{L,E}^{\psi,\rm in}+\psi_{R,E}^{\rm in}(x)\hat a_{R,E}^{\psi,\rm in}+\psi_{L,-E}^{\rm in}(x)\hat a_{L,E}^{\psi,\rm in\dagger}+\psi_{R,-E}^{\rm in}(x)\hat a_{R,E}^{\psi,\rm in\dagger}]\\
    &=\int_0^\infty\odif{E}\ab[\psi_{L,E}^{\rm out}(x)\hat a_{L,E}^{\psi,\rm out}+\psi_{R,E}^{\rm out}(x)\hat a_{R,E}^{\psi,\rm out}+\psi_{L,-E}^{\rm out}(x)\hat a_{L,E}^{\psi,\rm out\dagger}+\psi_{R,-E}^{\rm out}(x)\hat a_{R,E}^{\psi,\rm out\dagger}].
\end{align}
From Eqs.~\eqref{eq:dirac_left}, \eqref{eq:dirac_right}, \eqref{eq:dirac_left_g} and \eqref{eq:dirac_right_g}, the $S$-matrix is obtained as
\begin{equation}
    \begin{pmatrix}
        \psi_{L,E}^{\rm in}(x)\\
        \psi_{R,-E}^{\rm in}(x)
    \end{pmatrix}=
    \begin{pmatrix}
        -ie^{-i\frac{m^2-2E^2}{2g}}M_\psi&e^{-\frac{m^2\pi}{2g}}\\
        -e^{-\frac{m^2\pi}{2g}}&ie^{i\frac{m^2-2E^2}{2g}}M_\psi
    \end{pmatrix}
    \begin{pmatrix}
        \psi_{L,E}^{\rm out}(x)\\
        \psi_{R,-E}^{\rm out}(x)
    \end{pmatrix}.
\end{equation}
This leads to the following relation:
\begin{align}
    \hat a_{L,E}^{\psi,\rm out}&=-ie^{-i\frac{m^2-2E^2}{2g}}M_\psi \hat a_{L,E}^{\psi,\rm in}+e^{-\frac{m^2\pi}{2g}}\hat a_{R,E}^{\psi,\rm in\dagger}.\label{eq:bogoliubov_psi}
\end{align}

Similarly, the positive-energy solutions for the Klein-Gordon equation are normalized to satisfy the condition $|j_\phi(\pm\infty)|=1$ as follows:
\begin{align}
    \phi_{L,E}^{\rm in}(x)&\underset{x\to-\infty}{\simeq}e^{-i\frac{g}{2}x^2+iEx}\frac{1}{\sqrt{m}}\ab(-\frac{m}{2(gx-E)})^{-i\frac{m^2}{2g}+\frac12},\label{eq:kg_normalized_1}\\
    \phi_{L,E}^{\rm out}(x)&\underset{x\to-\infty}{\simeq}  e^{i\frac{g}{2}x^2-iEx}\frac{1}{\sqrt{m}}\ab(-\frac{m}{2(gx-E)})^{i\frac{m^2}{2g}+\frac12},\\
    \phi_{R,E}^{\rm in}(x)&\underset{x\to+\infty}{\simeq} e^{i\frac{g}{2}x^2+iEx}\frac{1}{\sqrt{m}}\ab(\frac{m}{2(gx+E)})^{i\frac{m^2}{2g}+\frac12},\\
    \phi_{R,E}^{\rm out}(x)&\underset{x\to+\infty}{\simeq} e^{-i\frac{g}{2}x^2-iEx}\frac{1}{\sqrt{m}}\ab(\frac{m}{2(gx+E)})^{-i\frac{m^2}{2g}+\frac12}.\label{eq:kg_normalized_4}
\end{align}
A field operator $\hat\Phi(x)$ can then be expanded by these functions as
\begin{align}
    \hat \Phi(x)&=\int_0^\infty\odif{E}\ab[\phi_{L,E}^{\rm in}(x)\hat a_{L,E}^{\phi,\rm in}+\phi_{R,E}^{\rm in}(x)\hat a_{R,E}^{\phi,\rm in}+\phi_{L,-E}^{\rm in}(x)\hat a_{L,E}^{\phi,\rm in\dagger}+\phi_{R,-E}^{\rm in}(x)\hat a_{R,E}^{\phi,\rm in\dagger}]\\
    &=\int_0^\infty\odif{E}\ab[\phi_{L,E}^{\rm out}(x)\hat a_{L,E}^{\phi,\rm out}+\phi_{R,E}^{\rm out}(x)\hat a_{R,E}^{\phi,\rm out}+\phi_{L,-E}^{\rm out}(x)\hat a_{L,E}^{\phi,\rm out\dagger}+\phi_{R,-E}^{\rm out}(x)\hat a_{R,E}^{\phi,\rm out\dagger}].
\end{align}
From Eqs.~\eqref{eq:kg_left}, \eqref{eq:kg_right}, \eqref{eq:kg_left_g} and \eqref{eq:kg_right_g}, the $S$-matrix is obtained as
\begin{equation}
    \begin{pmatrix}
        \phi_{L,E}^{\rm in}(x)\\
        \phi_{R,-E}^{\rm in}(x)
    \end{pmatrix}=
    \begin{pmatrix}
        ie^{-i\frac{m^2-2E^2}{2g}}M_\phi&ie^{-\frac{m^2\pi}{2g}}\\
        ie^{-\frac{m^2\pi}{2g}}&ie^{i\frac{m^2-2E^2}{2g}}M_\phi
    \end{pmatrix}
    \begin{pmatrix}
        \phi_{L,E}^{\rm out}(x)\\
        \phi_{R,-E}^{\rm out}(x)
    \end{pmatrix}.
\end{equation}
This leads to the following relation:
\begin{align}
    \hat a_{L,E}^{\phi,\rm out}&=ie^{-i\frac{m^2-2E^2}{2g}}M_\phi \hat a_{L,E}^{\phi,\rm in}+ie^{-\frac{m^2\pi}{2g}}\hat a_{R,E}^{\phi,\rm in\dagger}.\label{eq:bogoliubov_phi}
\end{align}
\end{document}